\documentclass[aps,pra,reprint,superscriptaddress,amsmath,amssymb,nofootinbib,floatfix]{revtex4-2}
\usepackage{graphicx}
\usepackage{bm}
\usepackage{physics}
\usepackage{braket}
\usepackage{dcolumn}
\usepackage{siunitx}
\usepackage[dvipsnames]{xcolor}
\usepackage{hyperref}
\hypersetup{hidelinks,pdftitle={Memory-kernel dynamics of a photonic memristor and steering in its coherent extension},pdfauthor={Karol Bartkiewicz, Patrycja Tulewicz, Adam Miranowicz}}

\begin{document}
\renewcommand{\topfraction}{0.9}
\renewcommand{\bottomfraction}{0.6}
\renewcommand{\dbltopfraction}{0.95}
\renewcommand{\textfraction}{0.05}
\renewcommand{\floatpagefraction}{0.8}
\renewcommand{\dblfloatpagefraction}{0.8}
\setcounter{topnumber}{3}
\setcounter{dbltopnumber}{3}
\setcounter{totalnumber}{5}

\title{Memory-kernel dynamics of a photonic memristor and steering in its coherent extension}
\author{Karol Bartkiewicz}
\email{bark@amu.edu.pl}
\affiliation{Institute of Spintronics and Quantum Information, Faculty of Physics and Astronomy, Adam Mickiewicz University, 61-614 Pozna\'n, Poland}
\author{Patrycja Tulewicz}
\affiliation{Institute of Spintronics and Quantum Information, Faculty of Physics and Astronomy, Adam Mickiewicz University, 61-614 Pozna\'n, Poland}
\affiliation{Poznan Supercomputing and Networking Center, Institute of Bioorganic Chemistry of the Polish Academy of Sciences, 61-704 Pozna\'n, Poland}
\author{Adam Miranowicz}
\affiliation{Institute of Spintronics and Quantum Information, Faculty of Physics and Astronomy, Adam Mickiewicz University, 61-614 Pozna\'n, Poland}
\date{\today}

\begin{abstract}
We study a photonic memristor in which single photons pass, one at a time, through a half-wave plate whose angle is updated by each detection event and relaxes between detections. Every photon is prepared afresh and destroyed at the detector, so the memory of the device is the angle of the plate, which is a classical variable. We show that the mean angle is given exactly by the detection rate filtered by an exponential memory kernel. For the output polarization state averaged over the detection records, we derive a Nakajima-Zwanzig equation and give its two-time kernel explicitly. This equation is exact if the plate does not relax, whereas with relaxation it is nonlinear and only approximate. Nevertheless, for moderate feedback jumps and a relaxation time much longer than the feedback response time, it reproduces the area of the hysteresis loop found in point-process simulations within $1.1\%$, while the mean-field law deviates by up to $109\%$. Unfortunately, the kernel equation fails for strong relaxation and weak coupling. In the mean-field limit, we obtain a damping rate increased by the feedback and a Debye law for the loop area. We also show, under assumptions stated in the paper, that the device is unsteerable and that its dependence on the feedback basis follows from Malus's law, i.e., classical polarized light behaves in the same way. Finally, we consider an extension in which the photons themselves are stored in delay loops, and we propose a steering test of entanglement storage together with its loss budgets. Under the stated assumptions, a violation of the proposed inequality shows that the photon pair is steerable and that the joint channel of the loops cannot be written as a mixture of product channels with an entanglement-breaking factor.
\end{abstract}

\maketitle

\section{Introduction}
\label{sec:intro}

Memristive devices are circuit elements whose resistance depends on the history of applied signals, used in neuromorphic computing~\cite{Yang2013}, in-memory computing and adaptive information processing. Chua~\cite{Chua1971} introduced the memristor as a fourth fundamental two-terminal circuit element, in which the voltage is proportional to the current through a memory-dependent resistance (memristance) and an internal state variable is driven by the current (Sec.~\ref{sec:classical}). Memristive behavior is characterized by a pinched hysteresis loop in the current-voltage plane, which reduces to a single-valued function at high driving frequencies~\cite{Chua2011,DiVentra2009}. Since their experimental realization at Hewlett-Packard Laboratories~\cite{Strukov2008}, memristors have been used in physical neural networks and non-volatile memory~\cite{Yang2013,DiVentra2009}.

Memristive behavior is hard to extend to the quantum domain, as unitary evolution has no natural memory, and the measurement required by any feedback-based memory usually destroys the coherent superpositions needed for quantum information processing~\cite{Nielsen2010,Wiseman2010}. Nevertheless, several quantum memristors have been proposed and realized. Pfeiffer \textit{et al.}~\cite{Pfeiffer2016} introduced the quantum memristor as a dissipative device whose decoherence is controlled by feedback from a continuous measurement and whose dynamics are inherently non-Markovian. This work was followed by implementations that use superconducting quasiparticle tunneling~\cite{Salmilehto2017}, a photonic beam splitter~\cite{Sanz2018}, digital gates~\cite{Guo2022} and frequency-entangled optical fields~\cite{GonzalezRaya2022}. Photonic experiments include the first demonstration with integrated waveguides and measurement-based feedback~\cite{Spagnolo2022}, neuromorphic computing on a photonic quantum memristor~\cite{Selimovic2025}, vacuum--one-photon qubits with a quantum-dot source~\cite{DiMicco2025}, entanglement and coherence dynamics~\cite{Ferrara2025}, and a pair of coupled integrated memristors with crossed feedback~\cite{Baldazzi2026}.

Photonic hardware is a promising platform for fast, energy-efficient neuromorphic and machine-learning accelerators~\cite{shastri2021photonics,tait2017neuromorphic,zhang2024optical}, many of which operate with $10^{6}$--$10^{7}$ photons per detection window, i.e., with classical waves. Here, we discuss what operationally distinguishes a quantum neuromorphic optical device from a classical one and what a single-photon memristor with feedback memory offers in this respect.

Memory effects in open quantum systems are described by non-Markovian dynamics~\cite{Breuer2002,Rivas2014}, e.g., by the Nakajima-Zwanzig equation~\cite{Nakajima1958,Zwanzig1960}, in which the evolution depends on the past through a memory kernel. In quantum memristors, the loop area was used as a memory quantifier~\cite{Salmilehto2017,GonzalezRaya2022}, and non-Markovianity was tuned in bosonic devices~\cite{Tang2023}. Exponentially filtered feedback was described by a Fokker--Planck master equation in Ref.~\cite{AnnbyAndersson2022}, where the fast filtered signal is removed by a Nakajima--Zwanzig projection in the Supplemental Material. A deterministic master equation can be used to treat general non-Markovian signal processing~\cite{deSousa2026}. There is also a proposal of memory sharing between photonic memristors, which is based on measurements~\cite{Lim2026}. Windowed memory rules are in use~\cite{Spagnolo2022,DiMicco2025,Selimovic2025}. The exponential moving average with solved exponential weights, which is the discrete form of our kernel with the gain tied to the decay, can be found in the Supplemental Material of Ref.~\cite{Selimovic2025}. A signed phase step per photon was analyzed as an Ornstein--Uhlenbeck lock of a fiber interferometer~\cite{Hacker2023}. The closed-form stiffness and phase variance of this lock give, at small jumps, the feedback contribution to the effective damping rate of our device and the length of the mean Bloch vector of its output photons (Sec.~\ref{sec:meanfield}). For such a lock with relaxation, we derive the Nakajima-Zwanzig equation for the record-averaged state and its hysteresis-loop law, together with their range under detection noise.

\begin{figure}[tbp]
\centering
\includegraphics[width=\columnwidth]{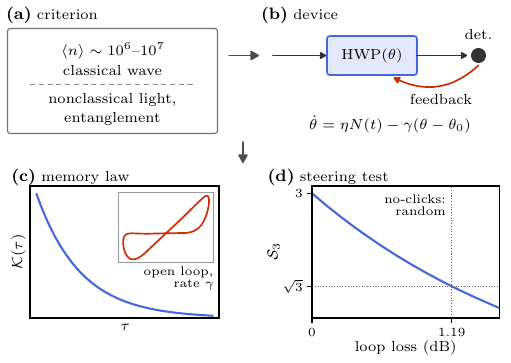}
\caption{Logic of the paper. (a)~Operational criteria separate quantum neuromorphic photonics (nonclassical light, entanglement) from classical-wave devices at $10^{6}$--$10^{7}$ photons per detection window. (b)~Analyzed single-photon memristor, whose half-wave-plate angle $\theta$, driven by detection-conditioned feedback and relaxation, is a classical register. (c)~Memory law of the mean register, with an exact exponential kernel $\mathcal{K}(\tau)$ of the time lag $\tau$ that decays at the relaxation (damping) rate $\gamma$, and its open-loop pinched loop (inset). Closing the feedback loop replaces $\gamma$ by the effective damping rate $\gamma_\text{eff}$ of small register deviations, which sets the loop area, and the output state obeys a Nakajima-Zwanzig equation (Fig.~\ref{fig:nz}). (d)~Coherent delay-loop extension. With three settings, ideal gates and unit efficiency outside the loops, the steering test holds up to $1.19$~dB per pass with random outcomes for no-clicks and $2.39$~dB with no-clicks as a third outcome. Note that a failed test does not show that the memory is classical.}
\label{fig:thesis}
\end{figure}

In this paper, we use linear-optical quantum information techniques~\cite{Lemr2015,Bartkiewicz2018} to develop a single model of a photonic memristor at the single-photon level, in which fresh photons drive a classical register. In Sec.~\ref{sec:theory}, we derive an exact exponential kernel for the mean register and, for the record-averaged output state, a Nakajima-Zwanzig equation with an explicit kernel and a tested range. The mean-field limit of this equation gives the Debye law of the hysteresis-loop area of the locked device. For a coherent delay-loop extension, we propose a steering test of entanglement storage with explicit loss budgets (Sec.~\ref{sec:fully_quantum}). Under stated assumptions, the feedback device is unsteerable and its coherent extension can pass this test (Fig.~\ref{fig:thesis}).

In contrast to path-encoded~\cite{Spagnolo2022} or Fock-encoded~\cite{DiMicco2025} implementations, we use polarization-encoded qubits and a half-wave plate (HWP) as the memory-dependent element. We keep the measurement-feedback mechanism of the photonic experiments of Refs.~\cite{Spagnolo2022,DiMicco2025,Selimovic2025} and analyze the closed loop beyond mean field. Since every photon is detected, our device, unlike that of Ref.~\cite{Spagnolo2022}, has no coherent output, but in both devices the memory is a classical register held in the feedback electronics. The superconducting approach~\cite{Salmilehto2017} requires millikelvin temperatures, and in gate-based implementations~\cite{Guo2022} the memristive dynamics are encoded in auxiliary qubits that emulate a collisional (repeated-interaction) memory. Another photonic route is provided by the proposals based on frequency-entangled optical fields~\cite{GonzalezRaya2022}. We compare these implementations with this work in Appendix~\ref{app:comparison}.\label{sec:comparison} We introduce a measurement-basis sensitivity index and show that the choice of feedback basis alters the hysteresis and that the same index can be obtained with classical polarized light.

In Secs.~\ref{sec:criteria} and \ref{sec:theory}, we discuss criteria for quantum neuromorphic hardware and derive the model with its Nakajima-Zwanzig equation and mean-field limit. We present our photonic implementation in Sec.~\ref{sec:implementation} and the basis-sensitivity index with numerical simulations in Sec.~\ref{sec:results}. Section~\ref{sec:fully_quantum} gives the steering test of the coherent delay-loop architecture and its loss budgets, and we conclude in Sec.~\ref{sec:conclusions}. In a companion paper~\cite{BartkiewiczPartII2026}, we develop networks of these memristors and investigate which of their functions meet the criteria of Sec.~\ref{sec:criteria}.

\section{Criteria for quantum neuromorphic hardware}
\label{sec:criteria}

A quantum neuromorphic optical system differs from its classical-wave analogue by two resources that can be tested separately~\cite{glauber1963coherent,mandel1995optical}. These are \emph{nonclassical light}, i.e., an input with no classical-wave description, such as a heralded single photon or sub-shot-noise (sub-Poissonian) light~\cite{glauber1963coherent,loudon2000quantum}, and \emph{entanglement}, which means that the dynamics generate non-separable multi-qubit states~\cite{werner1989quantum,aspect1982experimental}.

A coherent laser beam of power $P$ at wavelength $\lambda$ has a photon flux $\Phi=P\lambda/hc$, where $h$ is the Planck constant and $c$ the speed of light. For milliwatt powers at telecom wavelengths this flux is $\Phi\sim10^{15}$--$10^{16}\,$photons/s, e.g., $P=1$\,mW at $\lambda=1550$\,nm gives $\Phi\approx7.8\times10^{15}\,$photons/s. Thus, a nanosecond detection window contains of order $10^{6}$--$10^{7}$ photons (about $8\times10^{6}$ here), and analog optical networks operate with a mean photon number $\langle n\rangle\gg1$ per mode~\cite{zhang2024optical,tait2017neuromorphic,shastri2021photonics}. Here, $g^{(2)}(0)$ is the second-order correlation function at zero delay, i.e., the probability of detecting two photons at the same time normalized by the square of the single-photon detection probability. Note that attenuated laser light keeps $g^{(2)}(0)=1$ and squeezed states are nonclassical at $\langle n\rangle\gg1$~\cite{Paparelle2026}, so a low photon number does not suffice and is not required, but a measured $g^{(2)}(0)<1$ certifies nonclassical light~\cite{loudon2000quantum}. In Table~\ref{tab:criteria}, we compare representative platforms~\cite{li2021quantum,zhang2024optical,tait2017neuromorphic,hamerly2024quantum} by these criteria with order-of-magnitude photon numbers estimated from the flux relation, as they were not measured in every cited work.

\begin{table*}[tb]
\caption{Representative neuromorphic platforms against the criteria above. $\langle n\rangle$ is the order-of-magnitude photon number per mode/activation (our estimates from $\Phi=P\lambda/hc$ where not reported by the cited work).}
\label{tab:criteria}
\begin{ruledtabular}
\begin{tabular}{lcccc}
Platform & $\langle n\rangle$ & Entangl. & Sub-shot-noise & Class \\
\colrule
Superconducting qubits~\cite{li2021quantum} & $\lesssim1$ & available & n/a & [Q] \\
Microring / coherent-wave~\cite{zhang2024optical,tait2017neuromorphic} & $10^{6}$--$10^{7}$ & no & no & [C] \\
Few-photon classical light~\cite{hamerly2024quantum} & few & no & no & [C] \\
This work: feedback device & $<1$ & no & source only\footnotemark[1] & [C] memory \\
This work: coherent extension & $<1$ & via CPHASE & source only\footnotemark[1] & [Q] if certified \\
\end{tabular}
\end{ruledtabular}
\footnotetext[1]{Heralded spontaneous parametric down-conversion (SPDC) single photons are sub-Poissonian at the source, but we do not claim sub-shot-noise statistics at the device output. Detection backaction does not separate the feedback device from classical light, since attenuated light gives the same click statistics up to $g^{(2)}$. The [Q] entry of the coherent extension rests on the second criterion.}
\end{table*}

Standard certificates apply, e.g., an entangled pair is witnessed device-independently by a violation of the Clauser-Horne-Shimony-Holt (CHSH) Bell inequality~\cite{CHSH1969} between spacelike-separated measurements, and non-separability is detected by entanglement witnesses~\cite{guhne2009entanglement}. Below, we analyze a building block with a classical memory and a coherent delay-loop extension that can meet the second criterion in a certifiable way. The realistic loss up to which this certification holds is given in Sec.~\ref{sec:fully_quantum}.

\section{Theoretical framework}
\label{sec:theory}

In this section, we define the device and describe its dynamics at three levels. We start in Sec.~\ref{sec:model} with the device model and its correspondence with a classical memristor, while the polarization encoding and the dynamics photon by photon are specified in Sec.~\ref{sec:polarization}. Subsequently, we average over detection records and derive the Nakajima-Zwanzig equation for the output state (Sec.~\ref{sec:nonmarkov}). The mean-field limit of this equation, which we take in Sec.~\ref{sec:meanfield}, gives the effective damping rate and the closed-loop law of the loop area.

\subsection{Memristor and device model}
\label{sec:model}

\emph{Classical memristor.}---\label{sec:classical}A classical memristor is defined by a constitutive relation between the current $i(t)$ and the voltage $v(t)$ mediated by an internal state variable $w$~\cite{Chua1971,Chua2011,DiVentra2009}:
\begin{align}
v(t) &= M(w)\,i(t), \label{eq:memristor_iv}\\
\frac{dw}{dt} &= f(w, i(t)), \label{eq:memristor_state}
\end{align}
where $M(w)$ is the memristance (memory-dependent resistance) and the function $f$ describes how the current drives the state variable. A memristor is characterized by (i) a \textit{pinched hysteresis loop} passing through the origin, $i(v{=}0,t) = 0$ for all $t$, (ii) \textit{frequency dependence}, where, for a periodic drive of angular frequency $\omega$, the loop reduces to a single-valued function as $\omega \to \infty$, and (iii) \textit{non-volatile memory} in the ideal case, $dw/dt|_{i=0} = 0$.

\emph{Device.}---\label{sec:qm_def}We consider a device that performs a memory-dependent unitary transformation on each photon,
\begin{equation}
\ket{\psi_\text{out}(t)} = \hat{U}(\theta(t))\ket{\psi_\text{in}(t)},
\label{eq:qm_transform}
\end{equation}
where $\ket{\psi_\text{in}(t)}$ and $\ket{\psi_\text{out}(t)}$ are input and output states, and $\hat{U}(\theta(t))$ is a unitary parameterized by the classical memory variable $\theta(t)$, which evolves according to
\begin{equation}
\frac{d\theta}{dt} = \eta\, N(t) - \gamma(\theta - \theta_0),
\label{eq:theta_dynamics}
\end{equation}
where $N(t)$ is the point process of discrete photon detection events (a sum of Dirac deltas at detection times), $\eta$ is the feedback strength in radians per detection, and $\gamma$ is the relaxation (damping) rate of the register toward the equilibrium angle $\theta_0$, i.e., the inverse memory time. Hence, $\theta$ depends on the detection history, and integrating Eq.~\eqref{eq:theta_dynamics} over a single detection event yields a jump $\Delta\theta = \eta$. The formal solution of Eq.~\eqref{eq:theta_dynamics} reads
\begin{equation}
\theta(t) = \theta_0 + (\theta(0) - \theta_0)e^{-\gamma t} + \eta \int_0^t e^{-\gamma(t-t')} N(t')\,dt',
\label{eq:theta_solution}
\end{equation}
where the initial offset from $\theta_0$ decays at rate $\gamma$ and the last term accumulates the weighted detection history.

\emph{Correspondence with a memristor.}---\label{sec:correspondence}To establish the correspondence, we map the current $i(t)$ to the input photon rate $\Gamma_\text{ph}(t)$ and the voltage $v(t)$ to the output detection rate $R_\text{out}$. The internal state $w$ is mapped to the memory parameter $\theta$, while $P_H(\theta)$ corresponds to the memristance.

Firstly, the output detection rate,
\begin{equation}
R_\text{out} = \Gamma_\text{ph}\,P_H(\theta) = \Gamma_\text{ph}\cos^2(2\theta - \chi),
\label{eq:Rout}
\end{equation}
is proportional to the input photon rate $\Gamma_\text{ph}$ and vanishes for any memory state when no photons are present. Here $\chi$ parameterizes the input polarization $\ket{\psi_\text{in}} = \cos\chi\ket{H} + \sin\chi\ket{V}$ [Eq.~\eqref{eq:output_state} below], and $\Gamma_\text{ph}$ sets the conditional detection rate
\begin{equation}
R(t) = \mathbb{E}[N(t)\,|\,\theta(t)] = \Gamma_\text{ph}\,P_H(\theta),
\label{eq:detection_rate}
\end{equation}
which relates the point process $N(t)$ of Eq.~\eqref{eq:theta_dynamics} to its rate $R(t)$. Thus,
\begin{equation}
R_\text{out}\big|_{\Gamma_\text{ph}=0} = 0, \quad \forall\, \theta.
\label{eq:pinched}
\end{equation}
As for a classical memristor, where $v = M(w)\cdot i = 0$ when $i = 0$, the hysteresis loop passes through the origin for any memory state.

Secondly, for a prescribed mean detection rate (open loop), a periodic input $R(t) = R_0(1 + \sin\omega t)$ of mean $R_0$ and angular frequency $\omega$ yields a steady-state memory oscillation $\theta(t) = \theta_\text{DC} + \theta_\text{AC}\sin(\omega t - \phi)$ with the mean angle, amplitude and phase lag
\begin{equation}
\theta_\text{DC} = \theta_0 + \frac{\eta R_0}{\gamma},\quad
\theta_\text{AC} = \frac{\eta R_0}{\sqrt{\gamma^2 + \omega^2}},\quad
\phi = \arctan\frac{\omega}{\gamma}.
\label{eq:theta_oscillation}
\end{equation} The Lissajous loop area $A = \pi\theta_\text{AC}\sin\phi$ gives
\begin{equation}
A(\omega) \propto \frac{\eta\omega}{\gamma^2 + \omega^2},
\label{eq:loop_area}
\end{equation}
which vanishes in the quasistatic limit $\omega \to 0$ and for $\omega \to \infty$, where memory cannot follow, and is maximal at $\omega = \gamma$.

Finally, we note that the memory persistence can be tuned. In the absence of photon detections ($N = 0$), the memory parameter relaxes as $\theta(t) = \theta_0 + (\theta - \theta_0)e^{-\gamma t}$, so for $\gamma > 0$ the memory is volatile, in contrast to ideal classical memristors with $dw/dt|_{i=0} = 0$. Since $\gamma$ is programmed, $\gamma\to0$ can be used for non-volatile storage without light, but this fails under illumination at fixed $\eta$, where small perturbations of the register decay at the effective damping rate $\gamma_\text{eff}$. This rate is the sum of $\gamma$ and the restoring action of the feedback (Sec.~\ref{sec:kernel_strong}). The unipolar device, which jumps only on detections in the $\ket{H}$ channel, then locks where the probability $P_+$ of such a detection vanishes.

\subsection{Polarization qubits and discrete dynamics}
\label{sec:polarization}

For a photonic implementation using polarization-encoded qubits,
\begin{equation}
\ket{\psi} = \alpha\ket{H} + \beta\ket{V},
\label{eq:qubit}
\end{equation}
where $\ket{H}$ and $\ket{V}$ denote horizontal and vertical polarization and $|\alpha|^2 + |\beta|^2 = 1$, the memory-dependent transformation is realized by a half-wave plate at angle $\theta$ as
\begin{equation}
\hat{U}(\theta) = \begin{pmatrix}
\cos 2\theta & \sin 2\theta \\
\sin 2\theta & -\cos 2\theta
\end{pmatrix}.
\label{eq:hwp}
\end{equation}
Here, $\hat{U}(\theta)$ is Hermitian with $\hat{U}^2 = \openone$, i.e., a reflection of the real polarization plane. On the Bloch sphere, it is a rotation by $\pi$ about the axis $\vec{n} = (\sin 2\theta, 0, \cos 2\theta)$, which maps the Bloch vector $\vec{r}$ of the polarization state, with components $r_j=\langle\hat\sigma_j\rangle$, as
\begin{equation}
\vec{r} \to 2(\vec{n}\cdot\vec{r})\,\vec{n} - \vec{r},
\label{eq:bloch_reflection}
\end{equation}
where $\hat\sigma_{x,y,z}$ are the Pauli operators. The reflection axis is swept continuously with the memory parameter $\theta$. As an illustration, at $\theta = 0$ the plate applies $\hat{U}(0) = \hat{\sigma}_z$, which is a phase flip on $\ket{V}$, and $\hat{U}(\pi/8) = (\hat{\sigma}_x + \hat{\sigma}_z)/\sqrt{2}$ obtained at $\theta = \pi/8$ reflects through the diagonal axis. For $\theta = \pi/4$, we have $\hat{U}(\pi/4) = \hat{\sigma}_x$, which exchanges $\ket{H} \leftrightarrow \ket{V}$.

The output state for an input $\ket{\psi_\text{in}} = \cos\chi\ket{H} + \sin\chi\ket{V}$ is
\begin{equation}
\ket{\psi_\text{out}} = \cos(2\theta - \chi)\ket{H} + \sin(2\theta - \chi)\ket{V},
\label{eq:output_state}
\end{equation}
and the detection probability in the $\ket{H}$ channel is $P_H = \cos^2(2\theta - \chi)$, whose dependence on the accumulated memory $\theta$ is the source of hysteresis.

\emph{Discrete dynamics.}---\label{sec:master}The photons are processed sequentially, i.e., each photon $n$ is prepared in the same state $\rho_\text{in}=\ket{\psi_\text{in}}\!\bra{\psi_\text{in}}$, transformed by $\hat{U}(\theta_n)$ and destroyed at detection, after which $\theta$ is updated. Hence, the register $\theta$ is the only degree of freedom passed from photon to photon, and the photons are the ancillas of a collision model~\cite{Ciccarello2022}. The unconditional state of photon $n$ after the measurement is
\begin{equation}
\rho_n = \sum_j \hat{M}_j\,\hat{U}(\theta_n)\,\rho_\text{in}\,\hat{U}^\dagger(\theta_n)\,\hat{M}_j^\dagger,
\label{eq:discrete_map}
\end{equation}
where $\hat{M}_j$ are the measurement operators (e.g., $\hat{M}_H = \ket{H}\!\bra{H}$, $\hat{M}_V = \ket{V}\!\bra{V}$ for Z-basis measurement). Outcome $H$ occurs with probability
\begin{equation}
P_+(\theta_n)=\text{Tr}[\hat{\Pi}_+\hat{U}(\theta_n)\rho_\text{in}\hat{U}^\dagger(\theta_n)],
\label{eq:P_plus}
\end{equation}
where $\hat{\Pi}_+=\ket{H}\!\bra{H}$, so that for the pure input $P_+$ is the probability $P_H$ of Eq.~\eqref{eq:output_state}, and the register then obeys
\begin{equation}
\theta(t{+}\Delta t) = \theta(t) - \gamma(\theta - \theta_0)\Delta t + \eta\,\delta_{j,H},
\label{eq:conditional_feedback}
\end{equation}
with $\delta_{j,H}=1$ when the photon is detected in the $\ket{H}$ channel within a short time step $\Delta t$ and $0$ otherwise.

A photon leaving the plate at time $t$ is in the state
\begin{equation}
\rho(t)=\hat{U}(\theta(t))\,\rho_\text{in}\,\hat{U}^\dagger(\theta(t)),
\qquad
\vec{r}(t)=\big(\sin\varphi,\,0,\,\cos\varphi\big),
\label{eq:master}
\end{equation}
where $\vec{r}(t)$ is the Bloch vector of $\rho(t)$ and $\varphi=4\theta-2\chi$. For feedback in basis $B\in\{X,Y,Z\}$, each photon gives an outcome $m_B=\pm1$, and the mean signals are the components of $\vec{r}$,
\begin{equation}
\langle m_Z\rangle=\cos\varphi,\qquad\langle m_X\rangle=\sin\varphi,\qquad\langle m_Y\rangle=0.
\label{eq:signals}
\end{equation}
Since the measured photon does not return, measurement dephasing does not enter the loop, and different bases produce different trajectories. However, the signals in Eq.~\eqref{eq:signals} are normalized Stokes parameters, and classical polarized light with the same analyzers obeys the same mean-field laws.

\subsection{Nakajima-Zwanzig equation}
\label{sec:nonmarkov}

In this section, we derive a closed equation for the output state, which, unlike the mean memory angle, does not follow exactly from the detection rate. We first write the exact register kernel and the hierarchy of harmonics of the output phase. Subsequently, we eliminate the higher harmonics and compare the resulting equation with sampled detection records.

\subsubsection{Register kernel and harmonic hierarchy}
Since the register law is linear, by averaging Eq.~\eqref{eq:theta_dynamics} over detection records we obtain exactly
\begin{equation}
\langle\theta(t)\rangle = \theta_0 + \delta\theta_0\,e^{-\gamma t} + \int_0^t \mathcal{K}(t-s)\,\langle P_+\rangle(s)\,ds,
\label{eq:theta_formal}
\end{equation}
where $\delta\theta_0 = \theta(0) - \theta_0$ and
\begin{equation}
\mathcal{K}(\tau) = \eta\Gamma_\text{ph}\,e^{-\gamma\tau}
\label{eq:memory_kernel}
\end{equation}
is, at constant $\Gamma_\text{ph}$, the Green's function of the first-order filter programmed in the feedback law. This identity holds for any jump size, and point-process simulations at a jump of $0.5$~rad agree within their standard error. However, it is not a closed equation for the state, since $\langle P_+(\theta)\rangle\neq P_+(\langle\theta\rangle)$, and we derive the missing equation, exact at $\gamma=0$, below.

To derive it, we consider signed jumps, $+\Delta\theta$ on detector D$_1$ in the $\ket{H}$ channel and $-\Delta\theta$ on detector D$_2$ in the $\ket{V}$ channel (Fig.~\ref{fig:setup}), with $\eta\equiv\Delta\theta\,\Gamma_\text{ph}$ as in the signed laws of Sec.~\ref{sec:kernel_strong}. Let $c_n=\langle e^{in\varphi}\rangle$ be the record average, i.e., the average over detection records, with $c_0=1$. Then $c\equiv c_1=\langle r_z\rangle+i\langle r_x\rangle$ is the record-averaged output state. A jump multiplies $e^{in\varphi}$ by $e^{\pm ia_n}$, where $a_n=4n\Delta\theta$, with probabilities $(1\pm\cos\varphi)/2$, hence
\begin{equation}
\begin{split}
\partial_t c_n ={}& (A_n-in\nu)\,c_n+B_n\,(c_{n+1}+c_{n-1})\\
&-4in\gamma\,\langle(\theta-\langle\theta\rangle)\,e^{in\varphi}\rangle,
\end{split}
\label{eq:hierarchy}
\end{equation}
with $A_n=\Gamma_\text{ph}(\cos a_n-1)$, $B_n=\tfrac{i}{2}\Gamma_\text{ph}\sin a_n$ and $\nu=4\gamma(\langle\theta\rangle-\theta_0)+2\dot{\chi}$, so that $-\nu$ is the mean angular velocity of $\varphi$ between jumps. This hierarchy, with the chain structure of the Kuramoto-Daido harmonics~\cite{Tyulkina2018}, is the Fourier form of the Chapman-Kolmogorov equation of a first-order phase-locked loop with jumps~\cite{Forrester1973,Chie1978}, which in second order in $\Delta\theta$ gives the Fokker-Planck equation of the loop~\cite{Viterbi1963}. This hierarchy is exact and linear for any jump size at $\gamma=0$. For $\gamma>0$, we neglect the covariance in its last term, i.e., all records move with the mean velocity, and we close the system with $\langle\theta\rangle$ from $d\langle\theta\rangle/dt=\eta\,\text{Re}\,c-\gamma(\langle\theta\rangle-\theta_0)$.

\subsubsection{Equation and kernel}
\begin{figure*}[tbp]
\centering
\includegraphics[width=\textwidth]{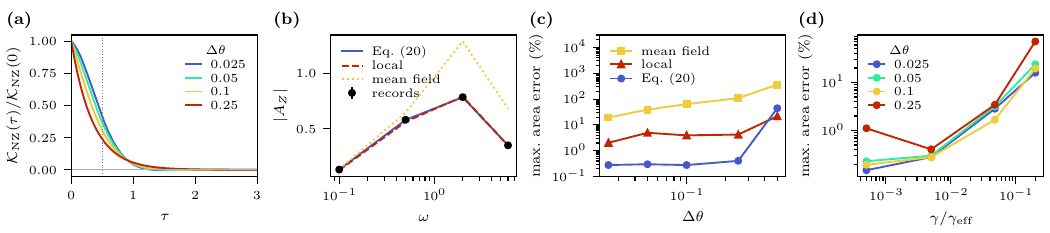}
\caption{Nakajima-Zwanzig equation against sampled detection records ($\eta=0.5$, signed jumps, $\chi=\tfrac12A_\chi\sin\omega t$, $240$ grid points with $\Delta\theta\in\{0.025,0.05,0.1,0.25,0.5\}$, $\gamma\in\{0.001,0.01,0.1,0.5\}$, $A_\chi\in\{0.2,1,2\}$, $\omega\in\{0.1,0.5,2,6\}$, where locked orbits have $A_\chi\omega<4\eta$). (a)~Kernel of Eq.~\eqref{eq:nz} at $\nu=0$ versus $\tau=t-s$, with $1/4\eta$ dotted. (b)~Loop area at $\Delta\theta=0.25$, $A_\chi=1$, $\gamma=0.01$ from the records, Eq.~\eqref{eq:nz}, the local closure and the scalar mean-field law. (c)~Largest area error on the grid at $\gamma=0.01$. (d)~Largest area error of Eq.~\eqref{eq:nz} versus the ratio $\gamma/\gamma_\text{eff}$ of the relaxation (damping) rate to the effective damping rate $\gamma_\text{eff}=\gamma+4\eta$ of the locked loop.}
\label{fig:nz}
\end{figure*}

Let $\mathcal{U}(t,s)$ propagate the block $n\geq2$ of Eq.~\eqref{eq:hierarchy}. Eliminating this block gives
\begin{equation}
\begin{split}
\partial_t c ={}& (A_1-i\nu)\,c+B_1[1+J(t)]\\
&+\int_0^t\mathcal{K}_\text{NZ}(t,s)\,c(s)\,ds,\\
\mathcal{K}_\text{NZ}(t,s) ={}& B_1B_2\,[\mathcal{U}(t,s)]_{22},
\end{split}
\label{eq:nz}
\end{equation}
where $[\mathcal{U}]_{nm}$ is the element of the propagator between harmonics $n$ and $m$, and $J(t)=\sum_{m\geq2}[\mathcal{U}(t,0)]_{2m}\,c_m(0)$ describes the initial state. For constant $\nu$, the Laplace transform of the kernel, with the Laplace variable $z$, can be written as the continued fraction~\cite{Mori1965}
\begin{equation}
\begin{split}
\tilde{\mathcal{K}}_\text{NZ}(z)&=B_1B_2\,g_2(z),\\
g_n(z)&=[z-A_n+in\nu-B_nB_{n+1}\,g_{n+1}(z)]^{-1}.
\end{split}
\label{eq:nz_continued_fraction}
\end{equation}
The kernel starts at
\begin{equation}
B_1B_2=-\tfrac14\Gamma_\text{ph}^2\sin(4\Delta\theta)\sin(8\Delta\theta),
\label{eq:nz_kernel_start}
\end{equation}
which tends to $-8\eta^2$ as $\Delta\theta\to0$, and has a width of order $1/4\eta$ for $\Delta\theta\leq0.25$ [Fig.~\ref{fig:nz}(a)]. The kernel depends on two times through $\nu$. If it is taken at the undriven state, the error in the loop area is up to $13\%$ at drive amplitudes $A_\chi\geq1$ of the input angle $\chi(t)=\tfrac12A_\chi\sin\omega t$.

For $\gamma=0$ and a prescribed $\chi(t)$, Eq.~\eqref{eq:nz} is the exact Nakajima-Zwanzig equation~\cite{Nakajima1958,Zwanzig1960,Breuer2002,Rivas2014} of the projection onto $c$. However, for $\gamma>0$, $\nu$ depends on $\langle\theta\rangle$, i.e., on the past of $c$, and $\mathcal{U}$ and $\mathcal{K}_\text{NZ}$ become functionals of $c$, so the equation keeps its form but is nonlinear and approximate. We observe that the kernel stays finite as $\Delta\theta\to0$, where the mean-field flow below is local in time. Therefore, it is a measure of the nonlinearity of the flow in the harmonics, and a memory integral is not sufficient for non-Markovian evolution~\cite{Mazzola2010}. With the kernel-free local closure $c_2\approx c^2|c|^2$ of a wrapped Gaussian phase, we obtain the loop area of the locked device within $0.5\%$ at $\Delta\theta\leq0.1$ and $\gamma\leq0.01$. We need the kernel at $\Delta\theta=0.25$ and on unlocked orbits. There, at $\gamma=0.01$, the errors of the local closure are up to $2$--$5\%$, depending on $\Delta\theta$, whereas the error of Eq.~\eqref{eq:nz} is at most $0.41\%$ [see Fig.~\ref{fig:nz}(b,c)]. This description is classical, because the elimination acts on the statistics of a register and a fresh photon (Sec.~\ref{sec:nonmarkov_char}), unlike in Ref.~\cite{AnnbyAndersson2022}, where the filtered signal is eliminated in the opposite regime of a fast register.

\subsubsection{Range of validity}
\label{sec:nz_range}We integrate Eq.~\eqref{eq:hierarchy} with $n\leq n_\text{max}=\lceil\pi/2\Delta\theta\rceil-1$ (Appendix~\ref{app:nz}), and the two-time integral of Eq.~\eqref{eq:nz} gives the same result. Compared with event-driven records at $\gamma=0.01$ and $\Delta\theta\leq0.25$, we find a loop-area error of at most $0.41\%$ on all orbits, whereas the error of the scalar mean-field law is up to $109\%$ [Fig.~\ref{fig:nz}(c)]. For one locked point at $\gamma=0.001$, $\Delta\theta=0.25$ we find $1.1\%$. Thus, the equation holds for $\gamma/\gamma_\text{eff}\lesssim0.01$, $\Delta\theta\leq0.25$ and signed jumps. The neglected covariance gives an area error with a median of about $2.5\,\Delta\theta\,\gamma/\gamma_\text{eff}$ at $\omega\gtrsim\gamma_\text{eff}$, which is three to twelve times larger for fast and strong drive.

Unfortunately, the equation fails outside this range. For example, its error is $3.5\%$ at $\gamma=0.1$ and $71\%$ at $\gamma=0.5$ [Fig.~\ref{fig:nz}(d)], and $44\%$ at $\Delta\theta=0.5$, $\gamma=0.01$. For weak coupling, $\eta/\gamma\leq0.1$, it can also reverse the orientation of the loop. Since neglecting the covariance breaks $\varphi=4\theta-2\chi$, $\langle\theta\rangle$ is a neutral direction of the hierarchy on the locking plateau, and the long-time state is unreliable for $\gamma\geq0.1$ at $\Delta\theta=0.25$ (at $\gamma=0.5$, $t\approx100$, $\langle\theta\rangle=0.07$ against $0.30$ in the records). Note that $n_\text{max}$ is a stability bound, and at $\Delta\theta\geq0.25$ its truncation error is not negligible (Appendix~\ref{app:nz}). A numerical scan, which is not a proof, shows that pairs of harmonics across $a_n=2\pi$ make the next truncation unstable for $\Delta\theta\lesssim0.35$, and the truncated chain has spurious slow modes (rate $0.27$ at $\Delta\theta=0.25$). For the unipolar variant, with $A_n=\tfrac12\Gamma_\text{ph}(e^{ia_n}-1)$ and $B_n=A_n/2$, the algebra is the same, but the accuracy is lower, with an area error at $\gamma=0.01$ of $1$--$7\%$ at $A_\chi=1$ and up to $49\%$ at $A_\chi=0.2$ near a dark branch.

\subsection{Mean-field limit and closed-loop response}
\label{sec:meanfield}

In this section, we give the response of the closed loop in closed form. To this end, we first take the mean-field limit of the register law and linearize it, which gives the effective damping rate. We then derive the spectral law of the loop area and its correction due to detection noise.

Here we replace $\langle P_+(\theta)\rangle$ by $P_+(\langle\theta\rangle)$, which for signed jumps is the closure $c_2\approx c^2$ with $|c|=1$ and gives Eq.~\eqref{eq:z_feedback}. This replacement neglects the fluctuations of the record and becomes exact over finite times for $\Gamma_\text{ph}\to\infty$ at fixed $\eta\Gamma_\text{ph}$. For the unipolar law of Eq.~\eqref{eq:conditional_feedback}, with $\eta$ the jump and $\theta$ written for $\langle\theta\rangle$, the state of Eq.~\eqref{eq:master} and the register form the closed system
\begin{equation}
\frac{d\theta}{dt} = \eta\,\Gamma_\text{ph}\,\text{Tr}[\hat{\Pi}_+\,\rho(t)] - \gamma(\theta - \theta_0).
\label{eq:theta_coupled}
\end{equation}
Now, the kernel of Eq.~\eqref{eq:memory_kernel} is only another way of writing $\theta(t)$. It is easy to see that, for a given $\chi(t)$, the law is local in time on a branch of the unwrapped $\varphi$, since a pure state fixes $\varphi=4\theta-2\chi$ up to $2\pi$, and that the history enters only through the branch number. However, the record-averaged state is mixed, and the closure holds only for weak mixing. In particular, for a signed jump $\Delta\theta$ at $\Delta\theta\,\Gamma_\text{ph}/\gamma=50$ (Sec.~\ref{sec:validation}), the phase variance $2\Delta\theta$ of Ref.~\cite{Hacker2023} gives $|\langle\vec{r}\rangle|\approx e^{-\Delta\theta}$ on the locked branch, which agrees within about $1\%$ at $0.1$~rad ($0.895$ against $0.905$), but not at larger jumps ($0.71$ against $0.78$ at $0.25$~rad).

\subsubsection{Linearized dynamics}
For constant $\chi$, the fixed point obeys
\begin{equation}
\theta_\text{ss} = \theta_0 + \frac{\eta\Gamma_\text{ph}}{\gamma}\,\text{Tr}[\hat{\Pi}_+\,\rho_\text{ss}],
\label{eq:theta_ss}
\end{equation}
with $\rho_\text{ss}=\hat{U}(\theta_\text{ss})\rho_\text{in}\hat{U}^\dagger(\theta_\text{ss})$. Since $\hat{U}(\theta)$ is a reflection whose axis rotates with $\theta$, we have $\mathcal{S}=\partial_\theta\rho|_{\theta_\text{ss}}=-2i[\hat{\sigma}_y,\rho_\text{ss}]$. An initial offset $\delta\theta(0)$ of the register displaces the output state by $\delta\rho(t)=\rho(t)-\rho_\text{ss}$, which relaxes as $\delta\rho(t)=\mathcal{S}\,\delta\theta(0)\,e^{-\gamma_\text{eff}t}$ with
\begin{equation}
\gamma_\text{eff} = \gamma-\eta\Gamma_\text{ph}\text{Tr}[\hat{\Pi}_+\mathcal{S}]=\gamma+2\eta\Gamma_\text{ph}\sin\varphi_\text{ss}.
\label{eq:gamma_eff_ss}
\end{equation}
where $\varphi_\text{ss}=4\theta_\text{ss}-2\chi$. Thus, $\gamma_\text{eff}$ is the effective damping rate of the closed loop, i.e., the rate at which a small displacement of the register from its fixed point, and with it the output state, decays when relaxation and feedback act together. It is the sum of the bare relaxation rate $\gamma$ and the feedback term $2\eta\Gamma_\text{ph}\sin\varphi_\text{ss}$, which is a restoring rate for $\sin\varphi_\text{ss}>0$. Note that the response relaxes at the rate $\gamma_\text{eff}$, although the register kernel has the width $\gamma$, and that the first-order loop is open at $\varphi_\text{ss}=0$.

\subsubsection{Closed-loop response}
\label{sec:nonmarkov_char}\label{sec:kernel_strong}The ratio $\eta\Gamma_\text{ph}/\gamma$ is the loop gain, i.e., the register is locked by the feedback for $\eta\Gamma_\text{ph}/\gamma \gg 1$, whereas for $\eta\Gamma_\text{ph}/\gamma \ll 1$ relaxation dominates. By memory we mean here that the output depends on the past input through the register. The register and the current photon form a Markov process and the output record is a hidden-Markov process, so its non-Markovianity is that of a classical signal~\cite{deSousa2026}, unlike that of a quantum channel acting on a persistent qubit.

We simulate the signed feedback variant with the mean-field law
\begin{equation}
\frac{d\theta}{dt} = \eta\,\langle m_Z\rangle - \gamma(\theta - \theta_0),
\label{eq:z_feedback}
\end{equation}
where in the signed laws $\eta\equiv\Delta\theta\,\Gamma_\text{ph}$ and $\chi(t)$ is the driven input polarization angle (Appendix~\ref{app:basis}). This law can be mapped exactly onto the unipolar law of Eq.~\eqref{eq:theta_coupled} by the substitutions $\eta\Gamma_\text{ph}\to2\eta$ and $\theta_0\to\theta_0-\eta/\gamma$. A small offset about the driven orbit $\theta^*(t)$ obeys
\begin{equation}
\frac{d\,\delta\theta}{dt} = -\gamma_\text{eff}(t)\,\delta\theta,
\qquad
\gamma_\text{eff}(t) = \gamma - \eta\,\frac{\partial\langle m_Z\rangle}{\partial\theta}\bigg|_{\theta^*(t)}.
\label{eq:gamma_eff_def}
\end{equation}
where $\gamma_\text{eff}(t)$ is the instantaneous effective damping rate of offsets from the driven orbit. For $\eta=0.5$, $\gamma=0.01$, the stable branch keeps $4\theta^* - 2\chi$ close to $\pi/2$. Thus, $\gamma_\text{eff}\approx\gamma + 4\eta$, and the decay has the cycle-averaged rate
\begin{equation}
\bar{\gamma}_\text{eff} = \gamma + 4\eta\left\langle \sin\!\big(4\theta^* - 2\chi\big)\right\rangle_\text{cycle}.
\label{eq:gamma_eff_bar}
\end{equation}
where $\langle\cdot\rangle_\text{cycle}$ denotes the average over one drive cycle. To test the linearization of this one-dimensional equation, whose prediction is the rate, we displace the memory by $\delta\theta_0=0.02$ at eight phases of the settled cycle. We obtain a single exponential with a rate agreeing with Eq.~\eqref{eq:gamma_eff_bar} to better than $10^{-3}$ for $\eta\in[0.05,0.8]$ (Fig.~\ref{fig:kernel_strong}), and at $\eta=0.5$ both are equal to $2.008$, which is two hundred times the bare $\gamma$. Experimentally, $\gamma_\text{eff}$ can be measured by offsetting the wave plate and recording the output relaxation. For Poisson arrivals at constant $\chi$, the mean offset decays over $t<1$ at about $1.96$, $1.8$, $1.4$ and $0.2$ for signed jumps of $0.025$, $0.1$, $0.25$ and $0.5$~rad, compared with $\gamma_\text{eff}=2.01$. Above $0.1$~rad, the decay is not a single exponential.

\subsubsection{Spectral density}
The Fourier transform of the memory kernel,
\begin{equation}
\tilde{\mathcal{K}}(\omega) = \int_0^\infty \mathcal{K}(\tau)\,e^{-i\omega\tau}\,d\tau = \frac{\eta\Gamma_\text{ph}}{\gamma + i\omega},
\label{eq:kernel_fourier}
\end{equation}
has a Lorentzian squared modulus of width $\gamma$. For a prescribed drive and small swings, the loop area follows it,
\begin{equation}
A(\omega) \propto -\text{Im}\,\tilde{\mathcal{K}}(\omega) = \frac{\eta\Gamma_\text{ph}\,\omega}{\gamma^2 + \omega^2},
\label{eq:area_spectral}
\end{equation}
which recovers Eq.~\eqref{eq:loop_area}. However, closing the loop changes the width, since on the locked branch of Eq.~\eqref{eq:z_feedback}, with $\theta=\pi/8+\delta$ up to a constant offset,
\begin{equation}
\cos(4\theta-2\chi)\approx2\chi-4\delta,\qquad
\dot{\delta}=-(\gamma+4\eta)\delta+2\eta\chi.
\label{eq:locked_linear}
\end{equation}
Hence, for small drive amplitude $A_\chi$, the $\langle\hat{\sigma}_z\rangle$--$\sin2\chi$ loop of Sec.~\ref{sec:basis_dependence} has the Debye form
\begin{equation}
A_Z(\omega) = \frac{4\pi\eta A_\chi^2\,\omega}{\gamma_\text{eff}^2 + \omega^2},
\qquad \gamma_\text{eff}=\gamma+4\eta.
\label{eq:area_closed}
\end{equation}
Numerically, Eq.~\eqref{eq:area_closed} holds within $1\%$ at $A_\chi=0.1$ for $\eta=0.05$ and $0.5$ and $\omega\in[0.1,10]\gamma_\text{eff}$, and at $A_\chi=1$ the amplitude decreases by $12$--$30\%$ with the peak near $\gamma_\text{eff}$. Detection noise modifies this law, and with $r$ the stationary $|c|$ of Eq.~\eqref{eq:hierarchy} on the locked branch, to first order in $\Delta\theta$
\begin{equation}
A_Z(\omega) = \frac{4\pi\eta A_\chi^2\,r^2\,\omega}{(\gamma+4\eta r)^2 + \omega^2},
\label{eq:area_dressed}
\end{equation}
so the stiffness $4\eta$ is reduced by $r$, which is $1$ in the lock of Ref.~\cite{Hacker2023}. At $\gamma=0$, the stationary harmonics $b_n=\langle e^{in(\varphi-\pi/2)}\rangle$ obey
\begin{equation}
b_{n-1}-b_{n+1}=2\tan(2n\Delta\theta)\,b_n,
\label{eq:bn_recursion}
\end{equation}
hence
\begin{equation}
r=1-\Delta\theta-\frac{\Delta\theta^2}{2}-\frac{\Delta\theta^3}{3}+\dots,
\label{eq:r_series}
\end{equation}
which up to $\Delta\theta^2$ is the Tikhonov law of a first-order loop~\cite{Viterbi1963}, and $r=0.975$, $0.895$ and $0.710$ at $0.025$, $0.1$ and $0.25$~rad. Compared with the exact linear response (a Markov chain at $\gamma=0$, event-driven records at $\gamma>0$) for $\omega\in[0.5,20]$ and $\gamma\leq0.01$, Eq.~\eqref{eq:area_dressed} holds within $0.3\%$ for $\Delta\theta\leq0.06$ and within $1\%$ for $\Delta\theta\leq0.15$ if $r$ is taken at the actual $\gamma$. It should be stressed that this formula is not exact beyond first order, as a second pole near $12\eta$ has weight $-\Delta\theta^2/2$, and a finite drive adds about $-1.5\,\Delta\theta A_\chi^2$ to the relative area at $\omega\geq2$. The peak is set by the bare $\gamma$ only for weak feedback on a linearly biased orbit ($\theta_0=\pi/8$), whereas at $\theta_0=0$ weak feedback drives $\theta$ at $2\omega$ and the peak moves to $\omega\approx\gamma/2$.

\section{Proposed experimental implementation}
\label{sec:implementation}

\begin{figure}[tbp]
\centering
\includegraphics[width=0.86\columnwidth]{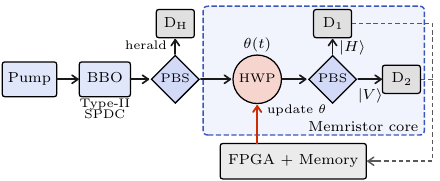}
\caption{Proposed experimental setup. Single photons from type-II spontaneous parametric down-conversion (SPDC), heralded by the idler detector D$_\text{H}$, are transformed by a feedback-controlled half-wave plate (HWP) whose angle $\theta(t)$ depends on the previous outcomes. A field-programmable gate array (FPGA) processes the outcomes at D$_1$ and D$_2$ and updates $\theta$ according to Eq.~\eqref{eq:conditional_feedback} or its signed variant ($\pm\Delta\theta$ on D$_1$ and D$_2$), and this closes the feedback loop.}
\label{fig:setup}
\end{figure}

\emph{Optical setup.}---\label{sec:setup}The proposed setup (Fig.~\ref{fig:setup}) consists of three stages, i.e., single-photon generation, memory-dependent transformation, and measurement with feedback.

Heralded single photons are generated by SPDC in a type-II beta-barium borate (BBO) crystal pumped at $\lambda_p = 405$~nm, producing signal and idler photons at $\lambda_{s,i} = 810$~nm with
\begin{equation}
\omega_p = \omega_s + \omega_i, \qquad \vec{k}_p = \vec{k}_s + \vec{k}_i.
\label{eq:phase_matching}
\end{equation}
where $\omega_{p,s,i}$ and $\vec{k}_{p,s,i}$ are the angular frequencies and wave vectors of the pump, signal, and idler photons. The idler photon detected at D$_\text{H}$ heralds the signal photon. The signal photon enters the memristor core, i.e., the feedback-controlled HWP implementing $\hat{U}(\theta)$ from Eq.~\eqref{eq:hwp} and a polarizing beam splitter (PBS) that routes photons to detectors D$_1$ ($\ket{H}$) and D$_2$ ($\ket{V}$), and the FPGA uses the outcome to update $\theta$.

\emph{Feedback system requirements.}---\label{sec:feedback}The feedback loop has to be completed before the next heralded photon arrives. As the updated wave-plate setting acts on subsequent photons, the loop need not be optically coherent with the detected photon, and its delay can be compared with the mean inter-photon spacing $\Delta t_\text{photons} \approx 100$~\textmu s at typical heralding rates of $10^{4}$~Hz. For superconducting-nanowire single-photon detectors (SNSPDs)~\cite{Natarajan2012} and, as in photonic feed-forward experiments~\cite{Prevedel2007}, a Pockels cell as the fast electro-optic actuator, the timing can be estimated as
\begin{equation}
\begin{aligned}
t_\text{detection} &\approx 10\text{ ns} &&\text{(SNSPD readout; jitter} < 100\text{ ps)},\\
t_\text{processing} &\approx 100\text{ ns} &&\text{(FPGA at 100 MHz)},\\
t_\text{actuation} &\approx 10\text{ ns} &&\text{(electro-optic rotator)},\\
t_\text{total} &\approx 120\text{ ns} &&\ll \Delta t_\text{photons} \approx 100\text{ \textmu s}.
\end{aligned}
\label{eq:timing}
\end{equation}

The feedback algorithm applies the jump at photon $n$, detected at $t_n$, and relaxes the setting continuously until the next photon as
\begin{equation}
\theta(t) = \theta_0 + \left(\theta(t_n) + \eta\,N_n - \theta_0\right)e^{-\gamma(t-t_n)},
\label{eq:discrete_feedback}
\end{equation}
where $N_n\in\{0,1\}$ records the detection in channel $H$.

The precision of the equivalent wave-plate angle has to be below $0.1^\circ$, which can be achieved with calibrated control of the electro-optic retardance. To keep the peak of the loop area within about $3\%$ (Sec.~\ref{sec:validation}), we need a signed jump $\Delta\theta\lesssim0.025$~rad. For $\gamma/\eta=0.02$ this jump gives $\gamma^{-1}\gtrsim2000$ photons, i.e., $\gtrsim200$~ms at $10^4$~Hz. For driving we also need modulators for $\chi(t)$ and $\Gamma_\text{ph}(t)$, and the register can be stored in the FPGA as a number, whose unwrapped value enters the relaxation and is applied modulo $\pi/2$ by the actuator (a rotator after a fixed half-wave plate).

For the test of Sec.~\ref{sec:fully_quantum} we need photon-number-resolving ancilla detection and fast in/out switches for the loops (their loss enters the arm efficiency $\eta_A$ of Sec.~\ref{sec:fq_budget}). We also need a tap-out splitter with a trusted tomography station for the feedback device, a quantum random-number generator fired after both photons leave the loops, and analyzer modulators fast enough for a new setting per clock cycle.

\section{Numerical results}
\label{sec:results}

The results on hysteresis and basis dependence follow from deterministic integration of the mean-field law (Appendices~\ref{app:iv} and \ref{app:basis}), and for the validation at the end of this section we simulated the stochastic detection process (Appendix~\ref{app:traj}). By default, we set $\eta = 0.5$, $\gamma = 0.01$, $\theta_0 = 0$, $\Gamma_\text{ph}=20$ (jump $\pm0.025$~rad) in the signed law and drive the input angle as $\chi(t)=(A_\chi/2)\sin\omega t$ with $A_\chi=1$, $\omega = 0.12$.

\emph{Hysteresis loops.}---\label{sec:hysteresis}\label{sec:frequency}For a prescribed signed drive $v(t)=\sin\omega t$ (open loop, no detection), the output $i(t)=G(\theta)\,v(t)$ with $G(\theta)=\sin^2(2\theta)$ traces a loop pinched at the origin, because $i\propto v$, and opened by the memory lag set by $\gamma$ (Eq.~\eqref{eq:theta_solution} and Fig.~\ref{fig:fingerprint}). This curve follows from the memory law alone, without photon-level simulation, and at high frequency, where the memory cannot follow the input, the loop collapses to $i=G(\theta_0)\,v$. Equation~\eqref{eq:loop_area} holds for swings $\theta_\text{AC}\ll\pi/4$, and at $\eta=0.15$ we find the peak of the area near $\omega=0.3$ instead of $\gamma=0.1$. With increasing $\eta$ the loop widens for $\theta_\text{AC}\lesssim0.5$~rad and narrows beyond this value. It also widens if we decrease $\gamma$. The photon-level pair $(\Gamma_\text{ph}(t),R_\text{out})$ of Sec.~\ref{sec:correspondence} passes through the origin when the input is modulated to zero, as for any lossy element. For weak modulation its area peaks at $\gamma_\text{eff}$ of Eq.~\eqref{eq:gamma_eff_ss}, which lies above $\gamma$ for $\sin\varphi_\text{ss}>0$ and below it for $\sin\varphi_\text{ss}<0$. The area vanishes as $\omega\to\infty$, whereas as $\omega\to0$ it vanishes only where the fixed point is unique, since near $P_+\approx0$ a fold gives static hysteresis.

\emph{Basis-sensitivity index.}---\label{sec:quantifiers}\label{sec:basis_sensitivity}We quantify the dependence of hysteresis on the feedback basis by the index
\begin{equation}
\mathcal{M}_{B_1,B_2} = \frac{|A_{B_1} - A_{B_2}|}{A_{B_1} + A_{B_2}},
\label{eq:basis_sensitivity}
\end{equation}
where $A_{B_i} = \oint \langle\hat{\sigma}_z\rangle_{B_i}\, dI$ is the area of one hysteresis cycle in the plane of the output $\langle\hat{\sigma}_z\rangle$ against the input observable $I(t)=\sin2\chi$ (Appendix~\ref{app:basis}) for feedback in basis $B_i$. An index value near unity indicates strong basis sensitivity. However, the index cannot indicate nonclassicality, as Eq.~\eqref{eq:signals} contains only normalized Stokes parameters and classical polarized light gives the same mean-field $\mathcal{M}$.

\emph{Basis-dependent hysteresis.}---\label{sec:basis_dependence}We consider the device with feedback, driven through the input angle $\chi(t)$, and study how the area of its Lissajous loop of $\langle\hat{\sigma}_z\rangle$, in general open, depends on the feedback basis $B$, whose outcomes are fed back into $\theta$. The feedback signals can be found in Eq.~\eqref{eq:signals}. Since the output is always observed in the Z basis and one photon cannot be measured twice, X and Y feedback requires a splitter.

\begin{figure}[tbp]
\centering
\includegraphics[width=\columnwidth]{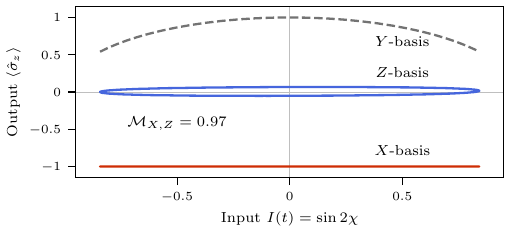}
\caption{Basis-dependent hysteresis from mean-field dynamics ($\eta = 0.5$, $\gamma = 0.01$, $\omega = 0.12$). Curves show the output observable $\langle\hat{\sigma}_z\rangle = \cos(4\theta_B - 2\chi)$ vs.\ input, with $\theta_B$ driven by basis-$B$ feedback. Z-basis feedback (blue) gives a thin hysteresis loop, whereas X-basis feedback (red) drives $\theta$ to a branch where $\cos(4\theta_X - 2\chi) \approx -1$ (nearly flat line). Y-basis feedback (grey, dashed) has zero mean signal, so at the mean-field level the response is the memoryless curve $\cos(2\chi)$. The index $\mathcal{M}_{X,Z} = 0.97$ quantifies this dependence.}
\label{fig:basis}
\end{figure}

In Fig.~\ref{fig:basis} we show the output observable $O=\langle\hat{\sigma}_z\rangle = \cos(4\theta - 2\chi)$ for three feedback bases. For feedback in the Z basis, $\theta$ is driven in phase with the detection probability and we obtain a thin hysteresis loop ($A_Z = 0.16$ per cycle). The hysteresis is almost removed ($A_X = 0.0026$ per cycle) by X-basis feedback, which drives $\theta$ via $\sin(4\theta - 2\chi)$ and moves $\theta$ to a branch near $\pm\pi/4$, where $\cos(4\theta - 2\chi) \approx -1$. In the case of the Y basis the mean feedback signal is zero for real polarization states, so at the mean-field level $\theta$ relaxes to $\theta_0 = 0$ and the output reduces to the single-valued curve $\cos(2\chi)$. We obtain $\mathcal{M}_{X,Z} = 0.97$, close to $(1-\varepsilon)/(1+\varepsilon)$ with $\varepsilon=\pi\gamma/4\eta$ at $\theta_0=0$, because $A_X\approx\varepsilon A_Z$. The index vanishes when each feedback basis is read out in that same basis, and a classical polarized beam with the same analyzers also gives $0.97$. We have checked that a photon clock and Poisson arrivals give the same spread of $\theta$, $0.06$ at $\pm0.025$~rad and $0.7$ at $\pm0.25$~rad, where the closure fails.

\emph{Validation of the mean-field register law.}---\label{sec:validation}We solved the scalar law of Eq.~\eqref{eq:z_feedback} by an adaptive Runge--Kutta (RK45) method and compared it with averages over $2\times10^{5}$ detection records sampled at the default parameters (Appendix~\ref{app:traj}), where each photon gives an outcome $\pm1$ with probability $(1\pm\cos\varphi)/2$ and a jump $\pm0.025$~rad. As the records hold no further quantum state, this tests the closure and the statistics of fresh photons. The root-mean-square (RMS) deviation of $\theta$ is $\sim 4\times 10^{-4}$, of the order of the Euler--RK45 difference ($3.5\times10^{-4}$), and its noise-induced part is of the same order.

For Poisson arrivals and fixed $\eta$, the RMS deviation of the mean $\theta$ stays below $2\times10^{-3}$, and at $\omega=0.12\ll\gamma_\text{eff}$ the loop area agrees within $1.5\%$ for jumps up to $0.1$~rad. However, this low-frequency area is insensitive to the width, as $\gamma_\text{eff}$ cancels in it. At the Debye peak ($\omega=2$) the area decreases by $3\%$, $12$--$15\%$, and $33$--$39\%$ for jumps of $0.025$, $0.1$, and $0.25$~rad, respectively ($A_\chi=0.3$--$1$), as described by Eq.~\eqref{eq:nz}, and for $0.5$~rad the area at $\omega=0.12$ decreases by $45\%$. The closure is worse where $P_+$ is small and requires $\Delta\theta\ll\tfrac12\sqrt{P_+}$, similarly to the unipolar law near a dark branch (Sec.~\ref{sec:nz_range}). For weak unipolar feedback, $\eta\Gamma_\text{ph}/\gamma\leq1$ with jumps up to $0.1$~rad, the rate of Eq.~\eqref{eq:gamma_eff_ss} reproduces the averaged kick response of $\theta$, although up to $10\%$ too fast. For the other bases the agreement is worse, e.g., for X-basis feedback the ensemble started at $\theta=0$ splits into two branches (RMS of the mean $\theta$ is $0.60$), whereas a start on one branch gives $6\times10^{-4}$. The Y curve of Fig.~\ref{fig:basis} is valid only at the mean-field level, because for Y-basis feedback the jumps diffuse $\theta$ and the mean output departs from $\cos2\chi$ with RMS of about $0.7$ at $0.025$~rad and $0.24$ at $0.0025$~rad.

\section{A steering test of entanglement storage}
\label{sec:fully_quantum}

In the device analyzed so far, successive photons interact only through the classical register $\theta$. Thus, their joint single-qubit statistics can be reproduced by a local classical model, and the index $\mathcal{M}_{B_1,B_2}$, which can be reproduced by a classical polarimetric loop, cannot exclude local hidden-variable models. We now replace the classical feedback loop with coherent photon storage in a three-qubit programmable controlled-phase gate with two delay loops. Before being measured, each photon enters the gate as the target and returns after a one-step delay as the program, and, two steps after entering, it serves as the control. The gate at time step $j$ therefore acts jointly on the three consecutive photons $(j, j{-}1, j{-}2)$ and applies pairwise controlled phases $\phi_1$ on the target--program pair and $\phi_2$ on the program--control pair, but no direct target--control phase. Since no measurement takes place inside the loops, coherence is preserved and each photon becomes directly entangled with its nearest neighbors $k\pm1$ and correlated with more distant photons through the resulting chain, as in delay-loop cluster-state sources~\cite{Pilnyak2017,Istrati2020}. The gates are diagonal and act between neighbors, so each output marginal depends only on inputs $k{-}1,k,k{+}1$. For product inputs $\cos\chi_k\ket{H}+\sin\chi_k\ket{V}$, the coherence of photon $k$ is
\begin{equation}
\begin{split}
&\sin2\chi_k\,f(\chi_{k-1})f(\chi_{k+1}),\\
&f(\chi)=\cos^2\chi+e^{-i(\phi_1+\phi_2)}\sin^2\chi,
\end{split}
\label{eq:fq_coherence}
\end{equation}
so a time-symmetric drive gives no loop area unless loss or birefringence in the loops breaks this symmetry. The test below concerns the delay lines and the gate.

\subsection{Coherent architecture and steering test}
\label{sec:fq_architecture}

\begin{figure}[tbp]
\centering
\includegraphics[width=\columnwidth]{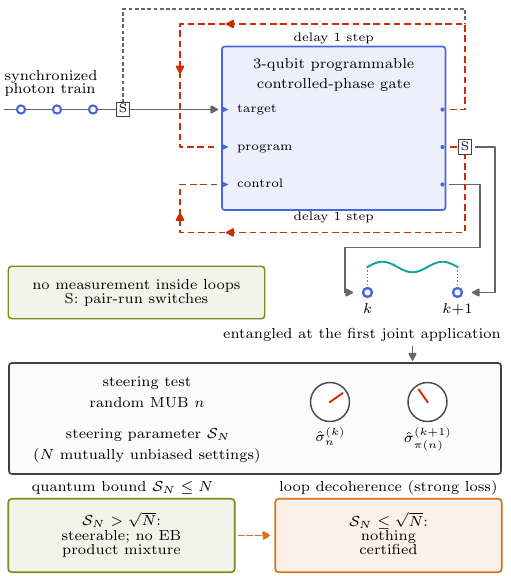}
\caption{Coherent delay-loop extension. Each photon of a synchronized train passes through a three-qubit programmable controlled-phase gate with two coherent delay loops as target, one step later as program, and two steps after entering as control. At time step $j$ the gate acts on the three consecutive photons $(j, j{-}1, j{-}2)$. Coherence is preserved, as no measurement takes place inside the loops. In the pair run, switches S send photon $k$ directly into the program-port delay and photon $k{+}1$ out to the analyzers. Successive output photons $k$ and $k{+}1$ are entangled at the first joint application when the second is idle. Measuring each in a randomly chosen mutually unbiased basis $\hat\sigma_n$ (with the later photon's settings relabeled in advance to match the steering map) and forming $\mathcal{S}_N$ [Eq.~\eqref{eq:temporal_steering}] can yield $\mathcal{S}_N>\sqrt{N}$ (up to the quantum bound $N$), a one-sided device-independent proof that the pair is steerable. With $\phi_1=\pi$, $\phi_2=0$, a trusted idle second application and no memory between passes, loops whose joint channel is a mixture of product channels with an entanglement-breaking (EB) factor obey $\mathcal{S}_N\le\sqrt{N}$.}
\label{fig:fqarch}
\end{figure}

To test entanglement storage, we use an otherwise empty loop (in a continuous train of photons in the state $\ket{+}=(\ket{H}+\ket{V})/\sqrt{2}$ the reduced pair is separable), where photons $k$ and $k{+}1$ share entanglement generated by the gate. Each is measured in one of $N$ mutually unbiased bases, chosen at random and labeled by its Pauli observable $\hat\sigma_n$ ($n=1,\dots,N$). We form the linear steering parameter~\cite{Cavalcanti2009,Saunders2010}, i.e., the correlator analog of the quadratic temporal-steering inequality of Refs.~\cite{PhysRevA.89.032112,bartkiewicz2016temporal,bartkiewicz2016experimental} with the same white-noise threshold,
\begin{equation}
\mathcal{S}_N=\sum_{n=1}^{N}\left\langle\hat\sigma_n^{(k)}\,\hat\sigma_{\pi(n)}^{(k+1)}\right\rangle.
\label{eq:temporal_steering}
\end{equation}
Here, the superscript labels the photon, and $\pi$ is a fixed relabeling chosen in advance and matched to the steering map. For the pair state of Eq.~\eqref{eq:pair_state}, $X$ on photon $k$ pairs with $Z$ on photon $k{+}1$ and vice versa, and $Y$ pairs with $Y$. With same-basis pairings the ideal correlators of this state would sum to exactly unity for every gate phase, and the bound could never be violated. A local-hidden-state model obeys $\mathcal{S}_N\le\sqrt{N}$, whereas a steerable pair can violate this bound, $\mathcal{S}_N>\sqrt{N}$, up to the quantum bound $\mathcal{S}_N\le N$ (Fig.~\ref{fig:fqarch}).

We assume that the measurement of photon $k{+}1$ is trusted and acts as characterized on a qubit, with loss independent of the settings and double clicks kept with a random outcome. Note that if double clicks were discarded, a product three-photon pulse could give $\mathcal{S}_3=3$ with no entanglement. We also assume that both settings are drawn at random after both photons leave the loops and that nothing depending on them, including the untrusted analyzer, affects photon $k{+}1$ before its measurement. Under these assumptions, a violation shows, in a one-sided device-independent way, that the pair is steerable from photon $k$. It should be stressed that the violation certifies the delay loops only with the gate phases fixed below, a trusted idle second application and no memory kept between the two passes through the loop crossed by both photons. In this case, the joint channel of the two loops is not a mixture of product channels with an entanglement-breaking factor~\cite{Pusey2015,Rosset2018}, and for independent noise neither loop is entanglement-breaking. However, collective noise can make each local channel fully depolarizing and leave the pair intact, and with such a memory one loop can entangle the pair by itself. As trusted-side loss is postselected, the test constrains transmission only along the arm of photon $k$. In the pair run both photons are measured together, so there is no channel between the two measurements and, in contrast to a temporal test in the sense of Ref.~\cite{PhysRevA.89.032112}, our scheme tests EPR steering of a time-ordered pair and entanglement storage.

We use steering, as it leaves the analyzer of photon $k$ untrusted and, with three settings, tolerates more white noise than a Bell test. With trusted analyzers the threshold is lower, and tests with a channel-threshold hierarchy or without trusted measurements can be found in Refs.~\cite{Ku2022,Graffitti2020,Rosset2018}. A nonzero generalized steering robustness $\mathcal{R}_T$, which is zero exactly for unsteerable assemblages (Sec.~\ref{sec:fq_budget} and Appendix~\ref{app:fq}), excludes the channels discussed above. As we show in Sec.~\ref{sec:dichotomy}, classical feedback memory yields $\mathcal{R}_T=0$, whereas for noiseless loops we have $\mathcal{R}_T=0.268$ at $N=3$.

The certification run has $n_g=2$ joint applications of the linear-optical controlled-phase gate and, after the first of them, one delay-loop traversal by each photon (Fig.~\ref{fig:fqarch}). The pair $(k,k{+}1)$ is injected into the empty loop, and photon $k$ is switched directly into the program-port delay bypassing a solo target pass, so it also crosses that loop before the first joint application. Both photons are switched out to the analyzers immediately after their second joint gate application, which means that photon $k{+}1$ passes the gate only twice and crosses one delay. We also assume that the two delay loops have equal per-traversal loss.

\begin{figure}[tbp]
\centering
\includegraphics[width=\columnwidth]{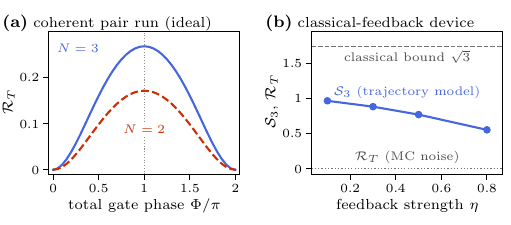}
\caption{Steering robustness of both devices. (a)~Coherent pair run (ideal hardware): $\mathcal{R}_T$ of the pair-run assemblage [Eq.~\eqref{eq:pair_state}] versus the total gate phase $\Phi$, for $N=3$ (solid) and $N=2$ (dashed) mutually unbiased settings, $\phi_2=0$. It is maximal at $\Phi=\pi$, where the pair is maximally entangled and the assemblage visibility is $\mathcal{V}=1$ exactly. (b)~Classical-feedback device in the tap-out protocol (trajectory model, $\gamma=0.01$, $\omega=0.12$): $\mathcal{S}_3$ stays well below the classical bound $\sqrt{3}$ (dashed) for all feedback strengths $\eta$, and $\mathcal{R}_T$ (dotted) is zero within the Monte Carlo noise floor, as required for a measure-and-prepare memory channel [Eq.~\eqref{eq:feedback_unsteerable}].}
\label{fig:devcert}
\end{figure}

In the pair run, each joint application involves the three-qubit programmable gate with one port in the vacuum, i.e., the control port in the first application (photon $k{+}1$ as target, photon $k$ as program) and the target port in the second (photon $k{+}1$ as program, photon $k$ as control). During certification the gate phases $\phi_1=\pi$, $\phi_2=0$ are classical settings of the gate and are not encoded in a program photon. As no train photon is detected inside the loop, ``heralded'' refers only to the ancilla detection pattern of the gate. As in the tunable linear-optical gates of Ref.~\cite{Lemr2015}, we model each application as a two-qubit controlled phase $\mathrm{CZ}(\phi_i)$ on the two occupied rails, where $\mathrm{CZ}(\phi)=\mathrm{diag}(1,1,1,e^{i\phi})$ in the basis $\{\ket{HH},\ket{HV},\ket{VH},\ket{VV}\}$. The heralding is preserved in this reduction, as the vacuum rail neither adds amplitude to the ancilla modes nor changes the accepted detection patterns. Within this gate model the two applications combine into a single controlled phase, and the ideal pre-measurement state of the pair can be expressed as
\begin{equation}
\ket{\psi(\Phi)} = \mathrm{CZ}(\Phi)\,\ket{+}\ket{+},
\qquad \Phi=\phi_1+\phi_2,
\label{eq:pair_state}
\end{equation}
which is maximally entangled at $\Phi=\pi$, as measuring photon $k$ in any basis leaves photon $k{+}1$ in a pure conditional state. Thus, the ideal assemblage visibility $\mathcal{V}=1$ is exact within this reduction. As shown in Fig.~\ref{fig:devcert}(a) for $\phi_2=0$, the robustness is maximal at $\Phi=\pi$, with the ideal-visibility values $\mathcal{R}_T=0.268$ ($N=3$) and $0.172$ ($N=2$). We test the loops only for this order of phases, because entanglement generated after the loops need not be stored. If the second phase is $\phi_2$, entanglement-breaking loops with honest analyzers give $\mathcal{S}_3=1+2\sin(\phi_2/2)$, which is $3$ at $\phi_2=\pi$ and $2.41$ at $\pi/2$, and exceeds $\sqrt{3}$ above $\phi_2=0.75$~rad. At $\phi_2=0$ they give $1$, and a fake analyzer sharing a hidden variable gives $\sqrt{3}$, i.e., the bound itself. Therefore, the gate phase has to be bounded independently, as for $\phi_2\neq0$ the fake analyzer exceeds the bound by about $\phi_2^2/(2\sqrt{3})$ at small $\phi_2$ and gives $1.90$ at $0.75$~rad.

\subsection{Loss budget under realistic imperfections}
\label{sec:fq_budget}

An imperfect gate contributes a process visibility $V_g$ per application, defined as the non-depolarized (white-noise-free) fraction of the three-qubit gate map. When photon $k$ clicks, the assemblage entering the steering test, i.e., the set of unnormalized states $\sigma_{a|x}$ of photon $k{+}1$ conditioned on outcome $a$ of setting $x$ on photon $k$, can be written as
\begin{equation}
\sigma_{a|x} = \tfrac{1}{2}\left[\, \mathcal{V}\,\Pi_{a|x} + (1-\mathcal{V})\,\tfrac{\openone}{2} \,\right],
\qquad
\mathcal{V} = V_g^{\,n_g},
\label{eq:budget_visibility}
\end{equation}
where $\Pi_{a|x}$ projects onto outcome $a$ of mutually unbiased setting $x$, up to a fixed relabeling of settings and outcomes under which $\mathcal{R}_T$ is invariant (e.g., for the state of Eq.~\eqref{eq:pair_state} an $X$ measurement on photon $k$ steers photon $k{+}1$ into $Z$ eigenstates). Since the pair-run loop is otherwise empty, a lost photon gives no click. Loss of the trusted photon $k{+}1$ is postselected, and trusted-side loss that depends on a hidden variable but on neither setting preserves the bound $\sqrt{N}$. However, loss of the untrusted photon $k$ cannot be postselected. A no-click does not show where the photon was lost, and the ancillas do not flag a missing photon. This photon crosses one loop before the first joint application and one after, so it clicks with probability $\epsilon=\eta_\text{loop}^{2}\eta_A$, where $\eta_\text{loop}$ is the transmission of one loop pass and $\eta_A$ collects the rest of its arm (switches, gate transits, detector) and the source, unless its failure is flagged before the setting is drawn.

If a random outcome is assigned to each no-click, we have $\mathcal{S}_N=N\mathcal{V}\epsilon$, which is to be compared with $\sqrt{N}$. By keeping the no-click as a third outcome~\cite{Bennet2012} and summing the correlators over clicks, we obtain $W_N=N\mathcal{V}\epsilon$, whereas a local hidden state that clicks at $m$ of the $N$ settings gives at most $\sqrt{m}$. Thus, at measured $\epsilon$ the bound is the concave envelope of $\sqrt{m}$ at $m=N\epsilon$, i.e.,
\begin{equation}
W_N\le1+(\sqrt{2}-1)(N\epsilon-1)\quad\text{for}\quad1\le N\epsilon\le2,
\label{eq:third_outcome_bound}
\end{equation}
and for $\mathcal{V}>1/\sqrt{2}$ a violation requires
\begin{equation}
\epsilon>\epsilon_N(\mathcal{V})=\frac{2-\sqrt{2}}{N(\mathcal{V}+1-\sqrt{2})},
\label{eq:epsilon_threshold}
\end{equation}
which is $1/N$ at $\mathcal{V}=1$. Below this visibility the next segment of the envelope applies, giving $\epsilon=1$ at $\mathcal{V}=1/\sqrt{N}$. The loss budgets for random outcomes and for a third outcome are, respectively,
\begin{equation}
\begin{split}
L_\text{dB} &< 5\log_{10}\!\left(V_g^{\,n_g}\eta_A\sqrt{N}\right),\\
L_\text{dB} &< 5\log_{10}\!\left[\eta_A/\epsilon_N(V_g^{\,n_g})\right],
\end{split}
\label{eq:loss_budget}
\end{equation}
where $L_\text{dB}$ is the per-pass loop loss in decibels. With independently measured parameters, we predict $\mathcal{S}_3=W_3=3V_g^{\,n_g}\eta_\text{loop}^{2}\eta_A$. A failed test does not show that the memory is classical, because depolarizing channels with $1/3<\mathcal{V}<1/\sqrt{3}$ are not entanglement-breaking and, with both analyzers trusted, they can be detected already by $\mathcal{S}_3>1$. These thresholds are specific to the two tests. In particular, loss before the gate leaves photon $k{+}1$ in $\ket{+}$ instead of a white state, and an assemblage-specific witness would tolerate more. For $N\in\{2,3\}$ mutually unbiased qubit settings (a qubit admits at most three), the white assemblage with visibility $\mathcal{V}$ is steerable if and only if $\mathcal{V}>1/\sqrt{N}$~\cite{CavalcantiSkrzypczyk2017}. We computed its generalized steering robustness~\cite{CavalcantiSkrzypczyk2017} $\mathcal{R}_T$ with the semidefinite program of Eq.~\eqref{eq:sdp_robustness} in Appendix~\ref{app:fq} (solved with \textsc{Mosek}), which also reproduces $\epsilon_N(\mathcal{V})$. In Fig.~\ref{fig:budget} we show $\mathcal{R}_T$ at $\mathcal{V}=V_g^{\,2}\eta_\text{loop}^{2}$ for $V_g\in\{1.00,\,0.95,\,0.90\}$, whose zero crossings give the first budget. The actual assemblage with random outcomes is not white and remains steerable up to $1.31$~dB at $V_g=1$, $N=3$. Assigning the standard threshold values~\cite{Bennet2012,CavalcantiSkrzypczyk2017} to this photon path, we obtain the budgets summarized in Table~\ref{tab:budget}.

\begin{table}[tbp]
\caption{Loss budget of the steering test, given as maximal per-pass loop loss [Eq.~\eqref{eq:loss_budget}], with photon $k$ crossing two loops, $n_g=2$ gate applications of visibility $V_g$, and $\eta_A=1$. Injecting photon $k$ at the program port without that first pass, from a simultaneous pair or a separate lossless delay, doubles the values.}
\label{tab:budget}
\begin{ruledtabular}
\begin{tabular}{lccc}
No-click rule & $V_g=1.00$ & $V_g=0.95$ & $V_g=0.90$ \\
\colrule
Random outcome, $N=3$ & $1.19$~dB & $0.97$~dB & $0.74$~dB \\
Random outcome, $N=2$ & $0.75$~dB & $0.53$~dB & $0.30$~dB \\
Third outcome, $N=3$ & $2.39$~dB & $1.99$~dB & $1.53$~dB \\
Third outcome, $N=2$ & $1.51$~dB & $1.11$~dB & $0.65$~dB \\
\end{tabular}
\end{ruledtabular}
\end{table}

For ideal gates and $N=3$ the test tolerates $1.19$~dB per loop pass with random outcomes and $2.39$~dB with a third outcome, and $0.74$ and $1.53$~dB at $V_g=0.90$, respectively. The three-qubit gate, itself a proposal, can be decomposed into two heralded two-qubit controlled-phase gates, so $V_g\approx V_{2q}^2$, where $V_{2q}$ is the visibility of one two-qubit gate, and both budgets vanish below $V_g=0.76$ ($V_{2q}=0.87$), which is about the value suggested by the conditional fidelities of Ref.~\cite{Lemr2015}. The storage loops of Refs.~\cite{PittmanFranson2002,Istrati2020}, without a gate, lost $0.92$ and $1.25$~dB per cycle and, at $V_g=0.90$, fit only the third-outcome budget. An arm efficiency $\eta_A=0.9$ lowers $1.19$ to $0.96$~dB and $2.39$ to $2.16$~dB, and an unheralded source with efficiency $\beta=0.7$ and detectors with efficiency $\eta_\text{det}=0.9$ ($\eta_A=0.63$) leaves $0.19$ and $1.38$~dB at $V_g=1$, and $0.53$~dB of the second budget at $V_g=0.90$. Note that the $100$~\textmu s photon spacing of Sec.~\ref{sec:feedback} applies only to the electronic loop of the feedback device. Here, the loop round trip has to match the train period, which requires a synchronized source at $10$--$100$~MHz and loops $3$--$30$~m long.

The gates have to be ancilla-heralded, as in Knill, Laflamme, and Milburn~\cite{KLM2001}, where a failure is flagged before the photon re-enters the loop. Unfortunately, postselected gates such as that of Ref.~\cite{Lemr2015} cannot be cascaded in the loop, and we use $V_g=0.90$ only as a representative visibility. If the ancilla detection resolves photon numbers with unit efficiency, heralding reduces the rate without lowering the visibility. We have found that the ancilla efficiency has to exceed about $92\%$, because at $90\%$ efficiency $15\%$ of accepted events per two-qubit gate carry a wrong photon number and, if they whiten the assemblage, the four two-qubit gates of a pair run give $\mathcal{V}\approx0.52<1/\sqrt{3}$. As only the two joint applications are heralded (Sec.~\ref{sec:fq_architecture}), for a success probability $p_{2q}\sim1/16$--$2/27$ of one heralded two-qubit gate the pair run succeeds with probability $p_{2q}^{4}\approx1.5$--$3\times10^{-5}$. With ideal photons this gives of order $10^{2}$--$3\times10^{3}$ pairs per second at $10$--$100$~MHz, times $(\beta\eta_\text{det})^{10}\approx10^{-2}$ for the ten photons at $\beta=0.7$, $\eta_\text{det}=0.9$. The longer run follows the port sequence of the architecture and is the harder test. If we omit the pre-gate pass and the idle application, we have $n_g=1$ and a plain controlled-phase gate with two delay lines, which requires no trust in the idle application and no memoryless loop. Its budgets are $1.93$ and $3.96$~dB at $V_g=0.90$, and $1.19$ and $2.48$~dB at $V_g=0.76$. To evaluate a finite sample, we propose a one-sided martingale test, as the raw $\mathcal{R}_T$ has a floor near $2.5/\sqrt{N_s}$ for $N_s$ samples per setting. In continuous operation, the pair can be isolated up to a known Pauli frame by $Z$ measurements on photons $k{-}1$ and $k{+}2$, provided the gates succeed without interruption.

\subsection{The feedback device is unsteerable}
\label{sec:dichotomy}
In the classical-feedback device the memory carried from $t_k$ to $t_{k+1}$ is the classical register $\theta$. For product input photons and the tester photon outside the loop, everything passing from one output photon to the next goes through a measure-and-prepare register, which is entanglement-breaking. Thus, if the test basis stays outside the feedback loop (as in the tap-out protocol below), the assemblage admits an explicit local-hidden-state model with the feedback trajectory $(\theta_k,\theta_{k+1})$ as the hidden variable, and
\begin{equation}
\mathcal{R}_T = 0,
\qquad
\mathcal{S}_N \le \sqrt{N},
\label{eq:feedback_unsteerable}
\end{equation}
for all $\eta$ and $\gamma$. Note that this result follows from the assumed classical register, so it shows what the device cannot do, but it does not measure its memory. We confirmed it numerically in the tap-out protocol, where every output photon is routed at a balanced splitter either to the device's fixed-basis feedback detector or to the tester, and the register relaxes per feedback detection event. For the assemblage of consecutive tester photons, sampled with the trusted state taken from the trajectory, we obtain $\mathcal{R}_T$ at the Monte Carlo noise floor ($\approx3\times10^{-3}$ at $N=3$), comparable to the residual signaling norm of the finite sample, up to which the steering test is meaningful. As the bound $\sqrt N$ holds for every fixed relabeling, we take the relabeling of Eq.~\eqref{eq:temporal_steering} to be the identity (same-basis pairings). For unipolar jumps $\eta\in[0.1,0.8]$~rad, we obtain $\mathcal{S}_3$ between $0.55$ and $0.965$, much lower than the bound $\sqrt{3}\approx1.73$ [Fig.~\ref{fig:devcert}(b)]. The steering test cannot be applied if the tester reconfigures the device's feedback measurement to the test basis $x$, because the basis choice then leaks into the memory and the resulting assemblage is signaling (marginal trace distance $0.32$ in the same model).

\section{Conclusions}
\label{sec:conclusions}

We have described a single-photon polarization memristor in which fresh photons drive a classical register. In this device, the input and output photon rates trace a loop through the origin which collapses at high frequency, and the persistence of its memory in the dark is programmable.

We have shown that the mean register obeys an exact exponential kernel (Eq.~\eqref{eq:memory_kernel}). For the record-averaged output state we derived the Nakajima-Zwanzig equation~\eqref{eq:nz} with an explicit two-time kernel, exact only at $\gamma=0$. Otherwise, this equation is a nonlinear approximation, which gives the loop area within $1.1\%$ of sampled records for jumps up to $0.25$~rad at $\gamma/\gamma_\text{eff}\leq0.005$. Unfortunately, it fails at strong relaxation, at weak coupling and near a dark branch (Sec.~\ref{sec:nz_range}). Note that its kernel describes classical statistics and that at small jumps a local closure works nearly as well. The mean-field limit is local in time on each branch of the register angle. By closing the loop, we renormalize the width from $\gamma$ to $\gamma_\text{eff}$, as in single-photon phase locks~\cite{Hacker2023}, which sets the loop area through the Debye law~\eqref{eq:area_closed}. The sampled records confirm the peak of the loop area within about $3\%$ at a signed jump of $0.025$~rad. However, at $0.1$~rad they give a peak lower by $12$--$15\%$, and on a nearly dark branch the closure fails earlier (Sec.~\ref{sec:validation}). For feedback in different bases we obtain $\mathcal{M}_{X,Z} = 0.97$, which is reproduced by classical polarized light, and under the stated assumptions the memory of the device is classical.

For the coherent delay-loop extension, we propose a steering test of entanglement storage at fixed gate phases $\phi_1=\pi$, $\phi_2=0$ and give its budgets in closed form (Eq.~\eqref{eq:loss_budget}, Table~\ref{tab:budget}). For three settings, ideal gates and unit efficiency outside the loops, the per-pass budget is $1.19$~dB with random outcomes and $2.39$~dB with no-clicks as a third outcome. Current gates stay below visibility $0.90$, at which these budgets become $0.74$ and $1.53$~dB. It should be stressed that the extension is a quantum memristor only in a restricted sense. Its output coherence [Eq.~\eqref{eq:fq_coherence}] vanishes with the input and is weighted by earlier inputs, which are stored as photons in the loops. However, this memory spans only two steps with a persistence that cannot be tuned, and for a time-symmetric drive we obtain no loop area. Note that the steering test certifies the quantum character of this memory, whereas a quantum memristor in the full sense requires a coherent memory with tunable persistence. Such persistence can be tuned by a variable beam splitter of amplitude reflectivity $r_\text{BS}$ in the loop, which keeps the photon for another round trip with amplitude $r_\text{BS}$, so that after $n$ round trips of duration $\tau_\text{loop}$ the stored amplitude is
\begin{equation}
r_\text{BS}^n=e^{-\gamma_q n\tau_\text{loop}/2},\qquad\gamma_q=-\frac{2\ln|r_\text{BS}|}{\tau_\text{loop}}.
\label{eq:coherent_kernel}
\end{equation} This is a coherent counterpart of the kernel of Eq.~\eqref{eq:memory_kernel}, with $\gamma_q$ in place of $\gamma$. However, each round trip adds loss, so the number of round trips over which the steering test still certifies the memory is bounded by the budgets of Eq.~\eqref{eq:loss_budget}.

The feedback device requires only commercially available components (heralded single-photon sources, fast feedback electronics and electro-optic polarization rotators), and its feedback loop is shorter than the time between photons ($t_\text{total} \approx 120\text{ ns} \ll \Delta t_\text{photons} \approx 100\text{ \textmu s}$). Recent neuromorphic reservoir computing with photonic quantum memristors~\cite{Selimovic2025} shows that such devices can be realized with current technology.

\section*{Acknowledgments}
We acknowledge funding from the European High Performance Computing Joint Undertaking (EuroHPC JU) under Horizon Europe Grant Agreement No.~101194322 (QEC4QEA), co-funded by the Polish National Centre for Research and Development (NCBiR) under decision No.~DWM/EuroHPC/2023/429/2025. The text was drafted and revised with the AI assistant Claude Code (Anthropic), and the authors take full responsibility for the article.

\section*{Data availability}
All simulation code and the data behind the figures and the numbers quoted in the text are openly available in the repository of Ref.~\cite{Repo2026}. No experimental data were generated in this work. Detailed numerical methods, derivations, and supporting figures are provided in the Appendices.

\appendix

\section{Numerical methods and reproducibility}
\label{app:methods}
All figures were produced with deterministic Python scripts (NumPy/SciPy, \textsc{cvxpy} with \textsc{Mosek} for semidefinite programs). In the case of stochastic figures we fixed the pseudo-random seed, so all reported numbers can be reproduced. The schematics in Figs.~\ref{fig:thesis} and \ref{fig:setup} are drawn programmatically, and Fig.~\ref{fig:thesis} uses the models of Appendices~\ref{app:iv} and \ref{app:fq}. If not stated otherwise, we integrate the leaky-integrator law of Eq.~\eqref{eq:theta_dynamics} for the memory by an adaptive Runge--Kutta method (relative tolerance $10^{-8}$) and take $G(\theta)=\sin^2(2\theta)$ as the power transmission of the half-wave plate. We used the AI assistant Claude Code (Anthropic; models Claude Sonnet 5.5 and Claude Opus 5.5, identifiers claude-sonnet-5-5 and claude-opus-5-5, in the final stage; model versions of earlier stages were not recorded) to write and run the code of the numerical simulations and figures, to derive and check analytical results, to check the numbers and the bibliography against the code and the sources, and to audit the results by independent reimplementation. We directed this work and checked the results by independent verification with separate programs and by manual consistency checks.

\section{Memristive fingerprint in i--V loops, time traces, and parameter sweeps}
\label{app:iv}
\begin{figure}[tbp]
\centering
\includegraphics[width=\columnwidth]{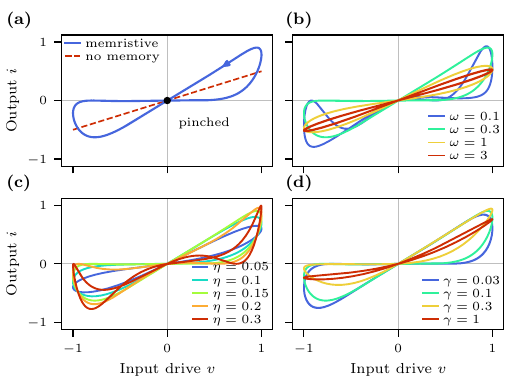}
\caption{Fingerprint of the register under a prescribed signed drive (open loop, no detection, $\eta=0.15$, $\gamma=0.1$, $\omega=0.3$, $\theta_0=\pi/8$). (a)~Pinched hysteresis loop. The output $i=G(\theta)\,v$ versus the signed drive $v$ passes through the origin, and the distinct branches reflect the memory in the HWP angle $\theta(t)$. The dashed red line is the memoryless ($\omega\!\to\!\infty$) response $i=G(\theta_0)\,v$. (b)~Frequency sweep ($\omega\in\{0.10,0.30,1.0,3.0\}$), where above $\omega\approx0.3$ the loop shrinks with $\omega$. At these parameters the swing reaches about $1$~rad and the peak lies near $\omega\approx0.3$ instead of at $\gamma$. (c)~Feedback-strength sweep ($\eta\in\{0.05,\dots,0.30\}$). The loop widens up to $\eta\approx0.15$ and narrows beyond. (d)~Relaxation-rate sweep ($\gamma\in\{0.03,\dots,1.0\}$), where smaller $\gamma$ gives more persistent memory and wider loops.}
\label{fig:fingerprint}
\end{figure}

\begin{figure}[tbp]
\centering
\includegraphics[width=\columnwidth]{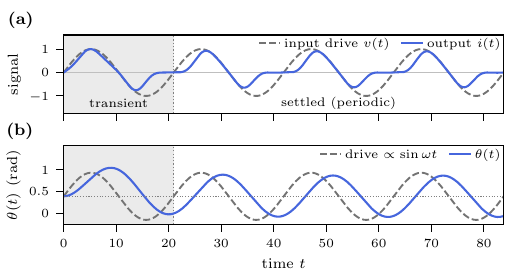}
\caption{Register evolution under a prescribed drive. (a)~Input drive $v(t)$ (dashed) and output response $i(t)=G(\theta)\,v$ (solid), whose memory-mediated phase lag produces the pinched hysteresis loop of Fig.~\ref{fig:fingerprint}(a). (b)~Memory parameter $\theta(t)$ (HWP angle) following the leaky integrator of Eq.~\eqref{eq:theta_dynamics}, which accumulates the drive and relaxes toward $\theta_0$ (dotted) with a lag. The shaded region, bounded by the dotted line at one drive period $T=2\pi/\omega$, marks the initial transient (memory relaxation on the scale $\sim 1/\gamma$) before the response settles into a limit cycle.}
\label{fig:evolution}
\end{figure}

We drive the memory with the signed signal $v(t)=V_0\sin\omega t$ according to
\begin{equation}
\dot\theta=\eta v-\gamma(\theta-\theta_0),
\label{eq:app_iv_feedback}
\end{equation}
where $i(t)=G(\theta(t))\,v(t)$ is the output modulated by the memory. The loop is pinched at the origin, since $i\propto v$, and it is opened into a figure-eight by the lag of the memory. For small swings the area of this loop follows $\eta\omega/(\gamma^2+\omega^2)$ [Eq.~\eqref{eq:area_spectral}]. At the nominal parameters $\eta=0.15$, $\gamma=0.10$, $\theta_0=\pi/8$, $\omega=0.30$, and $V_0=1$, this area has its peak near $\omega=0.3$. In our simulations we integrate eight to ten drive cycles and plot the last two or three cycles, which are settled. In Fig.~\ref{fig:evolution}(a) we show $v(t)$ and $i(t)$ over four cycles, and in Fig.~\ref{fig:evolution}(b) $\theta(t)$ against the scaled drive. In each sweep we vary one parameter, i.e., $\omega\in\{0.10,0.30,1.0,3.0\}$ in Fig.~\ref{fig:fingerprint}(b), $\eta\in\{0.05,0.10,0.15,0.20,0.30\}$ in Fig.~\ref{fig:fingerprint}(c), and $\gamma\in\{0.03,0.10,0.30,1.0\}$ in Fig.~\ref{fig:fingerprint}(d).

\section{Basis-dependent hysteresis}
\label{app:basis}
We make the input polarization angle of Eq.~\eqref{eq:output_state} sinusoidal in time, $\chi(t)=\tfrac{A_\chi}{2}\sin\omega t$, which defines the input observable $I(t)=\sin 2\chi$. The memory integrates a \emph{basis-dependent} feedback signal
\begin{equation}
\dot\theta=\eta\langle m_B\rangle-\gamma(\theta-\theta_0),
\label{eq:app_basis_feedback}
\end{equation}
with the signals of Eq.~\eqref{eq:signals}, and the plotted output is always the $Z$ observable $O=\cos(4\theta-2\chi)$ evaluated on the $\theta$ trajectory of the feedback basis $B$. For $\eta=0.5$, $\gamma=0.01$, $\theta_0=0$, $\omega=0.12$, and $A_\chi=1$, we integrate five cycles and retain the last two, quoting the areas per cycle. From the loop areas $A_B$, computed with the shoelace formula, we obtain the basis-sensitivity index $\mathcal{M}_{X,Z}=|A_X-A_Z|/(A_X+A_Z)=0.97$. A Jones-vector model of a bright polarized beam with analyzer feedback gives the same areas and index.

\section{Sampled detection records}
\label{app:traj}
\begin{figure}[tbp]
\centering
\includegraphics[width=\columnwidth]{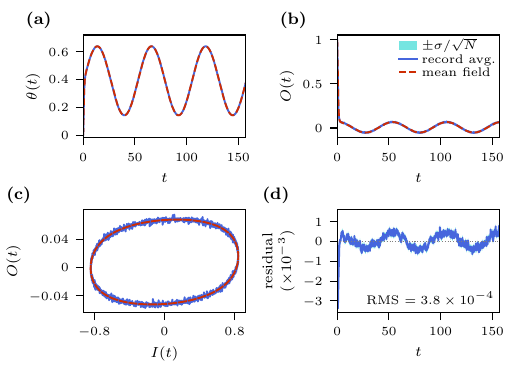}
\caption{Mean-field register law against sampled detection records ($\eta = 0.5$, jump $\pm0.025$~rad, $\gamma = 0.01$, $\omega = 0.12$, $2\times10^{5}$ records). (a)~Memory parameter $\theta(t)$. (b)~Output signal $O(t)$. (c)~Hysteresis loops of $O(t)$ versus the input observable $I(t)$ of Appendix~\ref{app:basis} (last two input cycles). (d)~Residual $\theta_\text{MF} - \langle\theta\rangle_\text{traj}$ with RMS error. Dashed red lines show the mean-field ODE, solid blue lines with shaded bands the record average $\pm 1\sigma/\sqrt{N}$. The RMS of $\sim\!4\times10^{-4}$ shrinks with the jump and does not decrease with $N$.}
\label{fig:validation}
\end{figure}

In Fig.~\ref{fig:validation} we compare Eq.~\eqref{eq:z_feedback}, integrated by RK45, with $N=2\times10^5$ records sampled at the parameters of the basis-dependent hysteresis ($\eta=0.5$, $\gamma=0.01$, $\omega=0.12$, $A_\chi=1$, step $\Delta t=0.05$). In each step we draw an outcome $m=\pm1$ with probability $(1\pm\cos(4\theta-2\chi))/2$ and change $\theta$ by $m\eta\Delta t=\pm0.025$~rad. The RMS deviation of $\approx3.8\times10^{-4}$ contains the Euler--RK45 difference and a noise-induced part, which both shrink with the jump, as here $\Delta t$ is the photon spacing. In the further tests of Sec.~\ref{sec:validation} we use Poisson arrivals at rate $\Gamma_\text{ph}=\eta/\Delta\theta$.

\section{Closed-loop response}
\label{app:kernel_strong}
We integrate the full nonlinear Z-feedback law $\dot\theta=\eta\cos(4\theta-2\chi)-\gamma(\theta-\theta_0)$ of Eq.~\eqref{eq:z_feedback}, used in Appendix~\ref{app:basis}, to its settled limit cycle $\theta^*(t)$ (six drive cycles, adaptive Runge--Kutta, relative tolerance $10^{-10}$). Next, we kick the memory, $\theta\to\theta+\delta\theta_0$ with $\delta\theta_0=0.02$, at eight equally spaced phases of the settled cycle and fit $|\delta\theta(t)|$ between kicked and unkicked solutions by a single exponential over an adaptive window of eight predicted decay times. The parameter-free prediction of Eq.~\eqref{eq:gamma_eff_bar} uses the same settled orbit. The measured and predicted rates, averaged over kick phases, agree to better than $10^{-3}$ for $\eta\in[0.05,0.8]$ ($\gamma=0.01$, $\omega=0.12$, $A_\chi=1$).

\begin{figure}[tbp]
\centering
\includegraphics[width=\columnwidth]{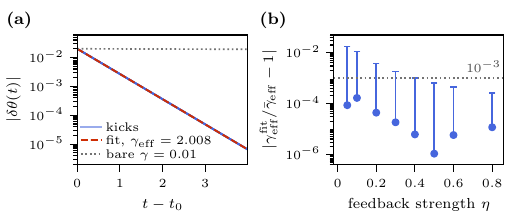}
\caption{Closed-loop response, with a feedback-renormalized width, of the scalar mean-field law ($\gamma=0.01$, $\omega=0.12$, Z-basis feedback, no linearization). (a)~Decay of the memory perturbation $|\delta\theta(t)|$ after a kick $\delta\theta_0=0.02$ at eight phases of the settled drive cycle for $\eta=0.5$. All traces collapse onto one exponential, whose fitted rate (dashed) matches the cycle-averaged linear-response prediction of Eq.~\eqref{eq:gamma_eff_bar}, and the bare damping rate $\gamma$ (dotted) is shown for comparison. (b)~Magnitude of the relative deviation of the fitted rate from the parameter-free prediction of Eq.~\eqref{eq:gamma_eff_bar} versus $\eta$ (upper error bars show the scatter over kick phases, logarithmic axis). All mean deviations lie at or below the $10^{-3}$ level (dotted), i.e., the fitted and predicted rates would be indistinguishable on a direct overlay.}
\label{fig:kernel_strong}
\end{figure}

\section{Nakajima-Zwanzig equation}
\label{app:nz}
As the reference we use an event-driven simulation with Poisson arrivals and exact relaxation of $\theta$ between arrivals, with no time step in the dynamics. We evaluate the area at the same times in the records and in all models, on one drive period after the hierarchy settles, with $\theta(0)=\pi/8$ and $\theta_0=0$. Note that $55$ points, $54$ of them at $\gamma\geq0.1$, do not settle within the cap and are compared at the same late times. Our estimator pairs each record at $\chi$ with one at $-\chi$ using common random numbers, and the standard error follows from the scatter of $24$ batches. On the $240$ grid points, the relative standard error of the area has median $0.098\%$ and $90$th percentile $0.20\%$. However, eleven points, all at $A_\chi=0.2$ and $\Delta\theta\geq0.25$, where the area is small, exceed $0.3\%$, so conclusions for them hold only in order of magnitude. At $\gamma=0$, where the hierarchy is exact, we find the residuals consistent with noise at $\Delta\theta\leq0.1$ (rms $z=0.8$--$1.25$, where $z$ is the residual in units of its standard error). With $n_\text{max}$ of the rule, we obtain rms $z$ of $1.5$ at $\Delta\theta=0.25$ and $12$ at $\Delta\theta=0.5$, and larger truncations ($n_\text{max}=40$ at $\Delta\theta=0.25$ and $n_\text{max}=12$--$40$ at $\Delta\theta=0.5$) give $1.0$ and $1.1$, respectively.

The hierarchy with $n\leq n_\text{max}$ is integrated by an eighth-order Runge-Kutta method (relative tolerance $10^{-8}$), where the local closure is the same system at $n_\text{max}=1$ with $c_2=c^2|c|^2$. We solve the two-time integral of Eq.~\eqref{eq:nz} by a pointwise exponential propagator on the block $n\geq2$ and the trapezoidal rule in the memory, iterating $\langle\theta\rangle$ within each step (at most four times). This scheme converges with order $2.00$ in the step and agrees with the hierarchy within $2.2\times10^{-5}$ at the time step $h=0.005$ ($8.9\times10^{-4}$ at $A_\chi=2$, $\omega=6$). Truncating the memory integral of Eq.~\eqref{eq:nz} at a cutoff time $\tau_c$, i.e., setting $\mathcal{K}_\text{NZ}(t,s)=0$ for $t-s>\tau_c$, with $\tau_c\geq4$ changes the area by less than $0.01\%$ at $\Delta\theta\leq0.25$. However, at $\Delta\theta=0.5$ a much longer memory is needed ($0.25\%$ at $\tau_c=32$), and too short a truncation breaks $|c|\leq1$ (e.g., $|c|=2.75$ at $\tau_c=0.1$).

The dropped covariance can be kept one level higher with $m_n=\langle(\theta-\theta_0)e^{in\varphi}\rangle$, $S=\langle(\theta-\theta_0)^2\rangle$, and the closure $\langle\delta\theta^2e^{in\varphi}\rangle\approx\sigma^2c_n$, where $\delta\theta=\theta-\langle\theta\rangle$ and $\sigma^2=\langle\delta\theta^2\rangle$ is the variance of the register. This closure does not cure the neutral direction of $\langle\theta\rangle$, although on the grid it gives an area error of at most $1.2\%$ for $\gamma\leq0.1$, $\Delta\theta\leq0.25$, compared to $3.5\%$ for Eq.~\eqref{eq:nz}. Note that the register may also be coherent, with each photon rotated by $\hat{U}(\hat\theta)$, where $\hat\theta$ is the angle operator of a quantum register, and the register then shifted by $\pm\Delta\theta$ conditioned on the photon's polarization. At $\gamma=0$ its populations obey the same Eqs.~\eqref{eq:hierarchy} and \eqref{eq:nz}, as the deferred-measurement principle requires, and the coherence appears only as entanglement of successive output photons, with negativity $\tfrac12\sin4\Delta\theta$ for a register in an eigenstate of $\hat\theta$.

\section{Comparison of quantum memristor implementations}
\label{app:comparison}
In Table~\ref{tab:comparison} we compare this work with the implementations of Sec.~\ref{sec:comparison}.

\begin{table*}[tbp]
\centering
\caption{Comparison of quantum memristor implementations identified by reference number. ``Memory'' is the physical mechanism supplying the memory, and ``Programmable'' indicates whether the memory response can be reshaped in situ via the feedback or control algorithm (``partial'' means tunable strength, fixed functional form). ``Memory law'' is the closed form given for the memory, and ``Coherent output'' tells whether a quantum input is mapped coherently to an output photon (``Partial'' means partially coherent). A dash denotes no or not applicable.}
\label{tab:comparison}
\begin{ruledtabular}
\begin{tabular}{lccccccc}
 & This work & Ref.~\cite{Spagnolo2022} & Ref.~\cite{Selimovic2025} & Ref.~\cite{DiMicco2025} & Ref.~\cite{Salmilehto2017} & Ref.~\cite{Guo2022} & Ref.~\cite{GonzalezRaya2022} \\
\colrule
Platform & Photonic & Photonic & Photonic & Photonic & Superconducting & Gate-based & Photonic \\
Encoding & Polarization & Path & Path & Fock & Flux & Qubit & Bosonic \\
Memory & Feedback & Feedback & Feedback & Feedback & Quasiparticles & Collisions & Feedback \\
Experiment & No & \checkmark & \checkmark & \checkmark & No & Simulation & No \\
Programmable & \checkmark & \checkmark & \checkmark & Partial & --- & \checkmark & Partial \\
Memory law & Exponential & Window & Window, exponential & Window & --- & --- & --- \\
Coherent output & No & \checkmark & Partial & \checkmark & --- & --- & --- \\
\end{tabular}
\end{ruledtabular}
\end{table*}

\section{Coherent extension and steering test}
\label{app:fq}
Figure~\ref{fig:fqarch} is a schematic (no numerical data) of the coherent delay-loop architecture and its steering test (see Eq.~\eqref{eq:temporal_steering} for the fixed, pre-declared setting relabeling).

\begin{figure}[tbp]
\centering
\includegraphics[width=\columnwidth]{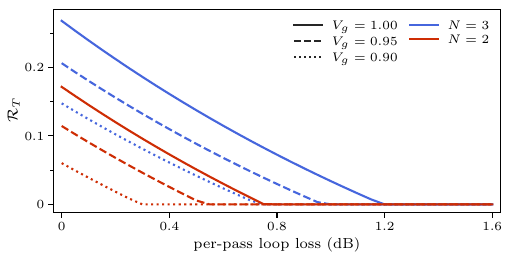}
\caption{Steering robustness $\mathcal{R}_T$ of a white assemblage computed by semidefinite programming (\textsc{Mosek}). We show $\mathcal{R}_T$ at visibility $\mathcal{V}=V_g^{\,2}\eta_\text{loop}^{2}$ versus per-pass loop loss for gate visibilities $V_g=1.00$, $0.95$, $0.90$ (solid, dashed, dotted) and $N=3$ (blue) or $N=2$ (red) mutually unbiased settings. The zero crossings reproduce the first budget of Eq.~\eqref{eq:loss_budget} and of Table~\ref{tab:budget}.}
\label{fig:budget}
\end{figure}

For Fig.~\ref{fig:budget} we consider the pair $\ket{\psi(\pi)}$ with isotropic noise of retained visibility $\mathcal{V}$, giving the assemblage of Eq.~\eqref{eq:budget_visibility} for $N$ mutually unbiased settings $\Pi_{a|x}$ on photon $k$. As a measure of quantum steering we use the generalized steering robustness~\cite{CavalcantiSkrzypczyk2017}, computed with the semidefinite program (SDP)
\begin{equation}
\begin{split}
\mathcal{R}_T=\min_{\{\sigma_\lambda\}}\ &\mathrm{Tr}\sum_\lambda\sigma_\lambda-1\\
\text{s.t.}\ \ &\sigma_\lambda\succeq0\quad\forall\lambda,\\
&\textstyle\sum_{\lambda:\lambda_x=a}\sigma_\lambda\succeq\sigma_{a|x}\quad\forall a,x,
\end{split}
\label{eq:sdp_robustness}
\end{equation}
which gives the minimal $t\ge0$ such that admixing weight $t/(1+t)$ of an arbitrary assemblage renders $\{\sigma_{a|x}\}$ unsteerable. Here, $x=1,\dots,N$ labels the setting, i.e., the mutually unbiased basis in which the untrusted photon $k$ is measured, and $a$ labels its outcome. The assemblage $\{\sigma_{a|x}\}$ is the set of unnormalized states of the trusted photon $k{+}1$ conditioned on outcome $a$ of setting $x$, $\sigma_{a|x}=p(a|x)\,\rho_{a|x}$, where $p(a|x)$ is the probability of outcome $a$ and $\rho_{a|x}$ the normalized conditional state. By no-signaling, $\sum_a\sigma_{a|x}$ is the reduced state of photon $k{+}1$ for every $x$. An assemblage is unsteerable if it admits a local-hidden-state model, $\sigma_{a|x}=\sum_\lambda\delta_{a,\lambda_x}\sigma_\lambda$ for all $a$ and $x$. Here, $\lambda=(\lambda_1,\dots,\lambda_N)$ runs over the $2^N$ deterministic strategies that assign the outcome $\lambda_x$ to setting $x$, $\delta_{a,\lambda_x}$ is the corresponding deterministic response function, and the local hidden states $\sigma_\lambda\succeq0$ of photon $k{+}1$ have unit total trace. In the SDP, the variables $\sigma_\lambda$ are such unnormalized hidden states without the trace condition, $A\succeq B$ means that $A-B$ is positive semidefinite, and $\mathrm{Tr}$ is the trace. Every feasible set $\{\sigma_\lambda\}$ gives $\sum_\lambda\delta_{a,\lambda_x}\sigma_\lambda=\sigma_{a|x}+t\,\tau_{a|x}$ with $t=\mathrm{Tr}\sum_\lambda\sigma_\lambda-1$ and an assemblage $\{\tau_{a|x}\}$, and division by $1+t$ yields an unsteerable assemblage. Hence, the minimum is $\mathcal{R}_T$, which vanishes if and only if $\{\sigma_{a|x}\}$ is unsteerable. We solve it with \textsc{Mosek} via \textsc{cvxpy} and recover the known $1/\sqrt N$ threshold visibilities. With the no-click as a third outcome, the same program over $3^N$ strategies reproduces $\epsilon_N(\mathcal{V})$ of Sec.~\ref{sec:fq_budget}. Figure~\ref{fig:budget} uses $\mathcal{V}=V_g^{2}\eta_\text{loop}^{2}$ and per-pass loss $L_\text{dB}=-10\log_{10}\eta_\text{loop}$.

Figure~\ref{fig:devcert} concerns both devices. For panel (a) we construct the pair-run state $\mathrm{CZ}(\Phi)\ket{+}\ket{+}$ exactly and evaluate its assemblage $\sigma_{a|x}=\mathrm{Tr}_A[(\Pi_{a|x}\otimes\openone)\rho]$ for $N=2,3$ mutually unbiased settings, and $\mathcal{R}_T(\Phi)$ follows from the same semidefinite program. By bisecting its zero crossing under white noise of visibility $V_g^{2}\eta_\text{loop}^{2}$, we reproduce the first budget of Eq.~\eqref{eq:loss_budget} to three decimals. The data of panel (b) come from the tap-out protocol, which we sample photon by photon from the discrete trajectory model ($4\times10^{5}$ photons, balanced routing, tester bases drawn uniformly at random). This gives $\mathcal{R}_T$ at the Monte Carlo noise floor and the reported $\mathcal{S}_3$ values. With the reconfigured-feedback control, where the basis choice enters the loop, we quantify the signaling ($0.32$) that voids the steering test.

\end{document}